\documentclass{article}

\usepackage{arxiv}

\usepackage[utf8]{inputenc} 
\usepackage[T1]{fontenc}    
\usepackage{hyperref}       
\usepackage{url}            
\usepackage{booktabs}       
\usepackage{amsfonts}       
\usepackage{nicefrac}       
\usepackage{microtype}      
\usepackage{lipsum}		
\usepackage{graphicx}
\usepackage{natbib}
\usepackage{doi}

\usepackage{amssymb}
\usepackage{latexsym}

\usepackage{url}
\usepackage{xcolor}
\usepackage{booktabs}
\usepackage{authblk}

\graphicspath{{./high-res-img/}}

\definecolor{newcolor}{rgb}{.8,.349,.1}

\title{DermX: an end-to-end framework for explainable automated dermatological diagnosis}%

\author[1,2]{Raluca Jalaboi}
\author[2]{Frederik Faye}
\author[2]{Mauricio Orbes-Arteaga}
\author[2]{Dan Jørgensen}
\author[1,3,4]{Ole Winther}
\author[2]{Alfiia Galimzianova}

\affil[1]{\footnotesize Department of Applied Mathematics and Computer Science at the Technical University of Denmark, Richard Petersens Plads, Building 324, DK-2800 Kongens Lyngby, Denmark}
\affil[2]{\footnotesize Omhu A/S, Silkegade 8 st, DK-1113 Copenhagen C, Denmark}
\affil[3]{\footnotesize Bioinformatics Centre, Department of Biology, University of Copenhagen, Copenhagen, Denmark}
\affil[4]{\footnotesize Center for Genomic Medicine, Rigshospitalet, Copenhagen University Hospital, Copenhagen, Denmark}
\date{}

\begin{document}
\maketitle

\begin{abstract}
Dermatological diagnosis automation is essential in addressing the high prevalence of skin diseases and critical shortage of dermatologists.
Despite approaching expert-level diagnosis performance, convolutional neural network (Conv\-Net) adoption in clinical practice is impeded by their limited explainability, and by subjective, expensive explainability validations.
We introduce DermX and DermX+, an end-to-end framework for explainable automated dermatological diagnosis.
DermX is a clinically-inspired explainable dermatological diagnosis Conv\-Net, trained using DermX\-DB, a 554 image dataset annotated by eight dermatologists with diagnoses, supporting explanations, and explanation attention maps. 
DermX+ extends DermX with guided attention training for explanation attention maps.
Both methods achieve near-expert diagnosis performance, with DermX, DermX+, and dermatologist F1 scores of 0.79, 0.79, and 0.87, respectively.
We assess the explanation performance in terms of identification and localization by comparing model-selected with dermatologist-selected explanations, and gradient-weighted class-activation maps with dermatologist explanation maps, respectively. 
DermX obtained an identification F1 score of 0.77, while DermX+ obtained 0.79.
The localization F1 score is 0.39 for DermX and 0.35 for DermX+.
These results show that explainability does not necessarily come at the expense of predictive power, as our high-performance models provide expert-inspired explanations for their diagnoses without lowering their diagnosis performance.
\end{abstract}


\section{Introduction}

\begin{figure*}[ht!]
\centering
\includegraphics[width=\textwidth]{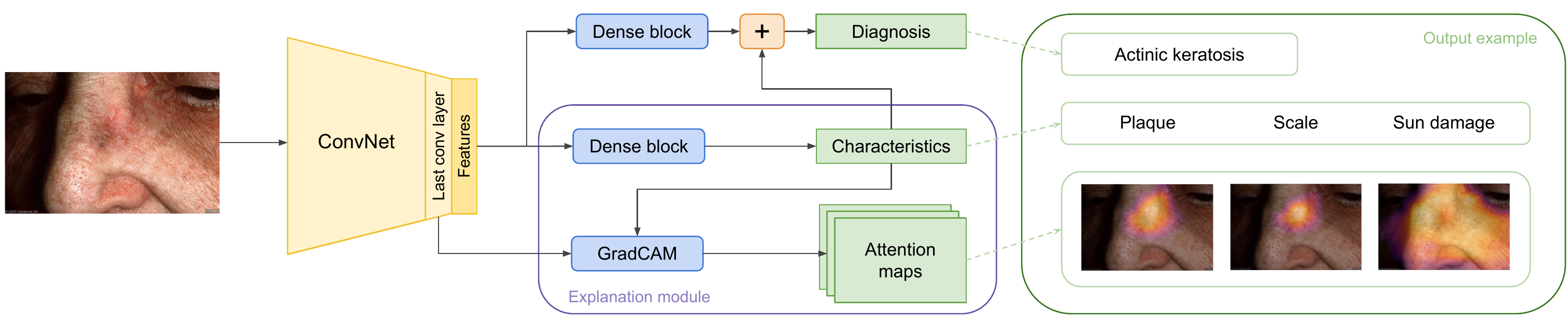}
 \caption{Clinically-inspired convolutional neural network architecture for image diagnosis with explanations in the form of skin lesion characteristics. 
 Given an image, the model is trained to predict the diagnosis together with the supporting characteristics, and to focus its attention on image sections that contain relevant characteristics.
 The diagnosis is predicted using both the characteristics identified by the model (similar to how dermatologists diagnose cases), and the extracted image features. 
 Using the extracted features alongside the predicted characteristics ensures that no relevant information is lost, e.g. the age or the skin tone.
 The explanation module offers plausible, faithful explanations to the diagnosis predicted by the model, while also localizing the explanations in the image.
 }
\label{fig:dermx_model_architecture}
\end{figure*}

\begin{figure}[ht!]
\centering
\includegraphics[width=0.65\linewidth]{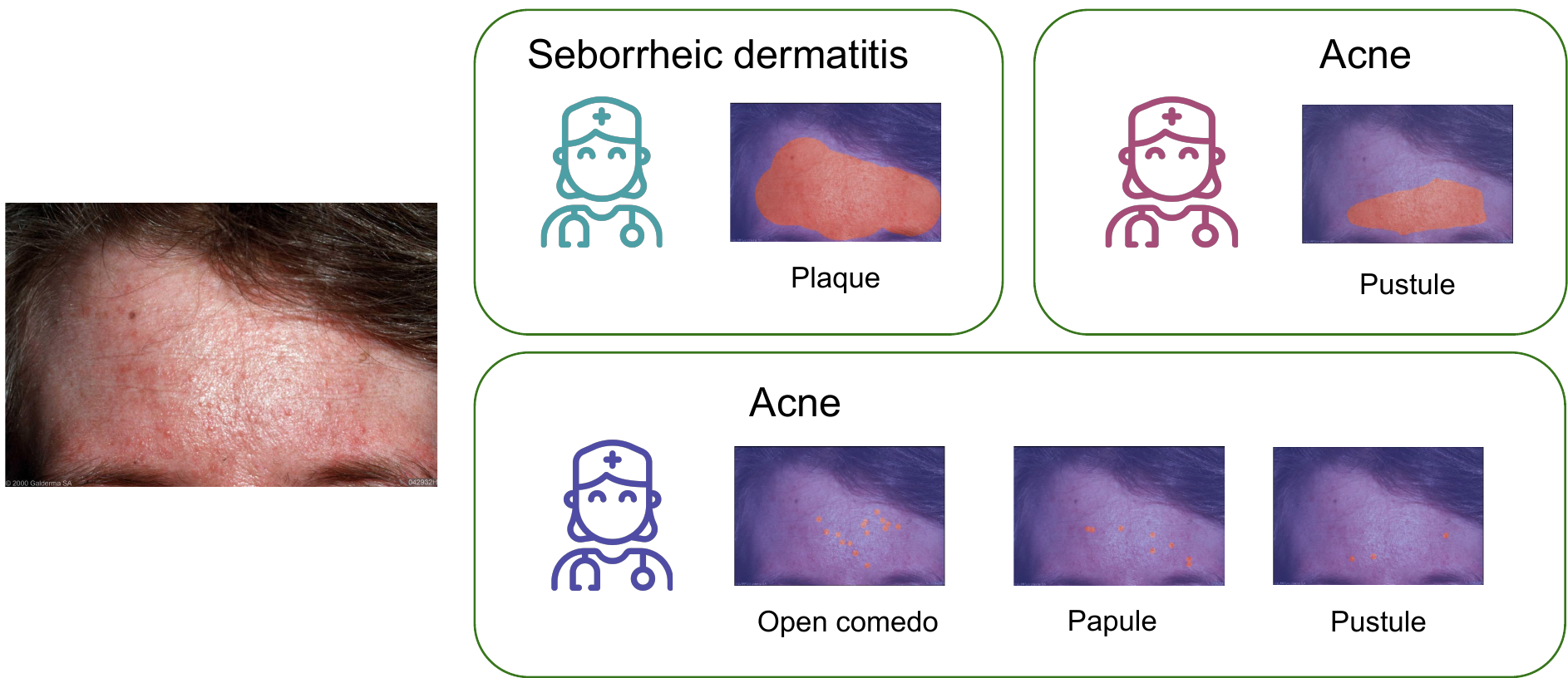}
 \caption{Sample from the DermX\-DB dataset. 
 A seborrheic dermatitis image from the SD-260 dataset was evaluated by eight dermatologists.
 Three evaluations are depicted in this figure. 
 One dermatologist correctly diagnosed it as seborrheic dermatitis due to the presence of plaque.
 Another dermatologist incorrectly diagnosed it as acne due to the presence of open comedones, papules, and pustules, while a third dermatologist diagnosed it as acne due to the presence of pustules.}
\label{fig:dataset_overview}
\end{figure}

Skin diseases affect a third of the global population~\citep{hay2014global} and are the fourth leading cause of disability worldwide~\citep{Karimkhani2017}.
The increasing demand for dermatological care is exacerbated by the low performance of general practitioners when diagnosing skin conditions~\citep{federman1999comparison}, and by the global scarcity of expert dermatologists~\citep{feng2018comparison, kringos2015building}. 

Automation may help alleviate this problem. 
Convolutional neural networks (Conv\-Nets) have been shown to achieve near expert-level performance in diagnosing dermatological conditions from images of skin lesions~\citep{thomsen2020systematic, esteva2017dermatologist}, and that they are able to assist general practitioners as well as less experienced dermatologists in improving their diagnostic performance~\citep{tschandl2020human, jain2021development}.
However, the lack of a good explanation mechanism~\citep{kelly2019key} for Conv\-Net decisions is one of the main obstacles to their adoption as automated diagnosis systems~\citep{goodman2017european, kelly2019key, topol2019high}.
A good explanation is expected to be both \emph{plausible},~i.e.~as similar as possible to a human explanation, and \emph{faithful},~i.e.~to accurately represent the inner workings of the network~\citep{jacovi2020towards}.

Different mechanisms for explaining Conv\-Net decisions have been proposed~\citep{simonyan2014deep, selvaraju2017grad, ribeiro2016should}.
Within the medical imaging literature, the most common explainability methods are saliency-based methods, such as raw saliency maps~\citep{simonyan2014deep} and gradient-weighted class-activation attention maps~(Grad-CAM)~\citep{singh2020explainable}.
While other methods were criticized due to their lack of faithfulness, Grad-CAMs have been shown to perform well~\citep{adebayo2018sanity}. 
However, there remains a lack of standard metrics for plausibility validation, as the explanations they provide are often incomplete and difficult to quantify~\citep{tschandl2020human}.
More specifically, common ConvNet explainability methods provide no semantic information alongside the explanation, but rather focus on the image section where the network pays attention.
In complex domains such as dermatology, this information is not enough to explain the decision mechanisms: knowing that the network focuses on the skin lesion does not explain why it diagnosed a case as acne and not rosacea.
Moreover, such complex tasks require that thorough explanation validation be done by domain experts, which is a time consuming and expensive process. 
Current dermatological datasets focus either solely on disease diagnosis, or on lesion segmentation~\citep{dermnetnz2021, sun2016benchmark, tschandl2018ham10000}. 
Having access to expert-annotated dermatological diagnosis explanations would improve the validation of explainability methods and allow the training of intrinsically explainable models.
However, to the best of our knowledge, no such dataset exists.

Our contributions are twofold.
First, to enable a quantitative assessment of the explainability of dermatological diagnosis models, we introduce DermX\-DB, a dermatological explainability dataset with gold standard diagnostic explanations provided by eight board-certified dermatologists. 
DermX\-DB consists of 554 images from DermNetNZ~\citep{dermnetnz2021} and SD-260~\citep{sun2016benchmark} associated with one of six diagnoses and their explanations in the form of skin lesion characteristics, as defined by~\cite{nast20162016}.
This labeling procedure mimics clinical practice, where dermatologists assess the characteristics of skin lesions to derive and support a tentative diagnosis~\citep{dermatology2017oakley}.
An annotation example can be seen in Figure~\ref{fig:dataset_overview}.

Second, we introduce DermX -- a novel, clinically-inspired Conv\-Net architecture for skin disease diagnosis and explanations. This architecture is illustrated in Figure~\ref{fig:dermx_model_architecture}.
Following the clinical approach of explaining dermatological diagnoses through skin lesion characteristics, DermX first identifies relevant characteristics in the image (which can also be interpreted as diagnosis explanations), and then relies on them, alongside the image features, to diagnose the case.
Using Grad-CAM~\citep{selvaraju2017grad}, we then localize the predicted characteristics in the image.
We validate the plausibility and faithfulness of our explanations using DermX\-DB as the gold standard for explanations~\footnote{The DermX\-DB dataset and the implementation of DermX and DermX+ are available at \url{https://github.com/ralucaj/dermx}}.

\subsection{Related work}
Machine learning-based dermatological diagnosis systems have been widely investigated, achieving results on par with human experts~\citep{esteva2017dermatologist, tschandl2020human, jain2021development}.
These advances in automated diagnosis of skin lesions were made possible in part by the emergence of various dermatological datasets, which contain images diagnosed by medical experts~\citep{tschandl2018ham10000, dermnetnz2021, sun2016benchmark}.
The widely used ISIC dataset~\citep{tschandl2018ham10000} also includes lesion segmentations that can partially serve as a basis for objective explanation measurement.
However, these segmentations were not collected to explain the diagnosis, but rather to localize the lesions. 
This shortcoming becomes critical in diseases such as actinic keratosis, where the area surrounding the lesion is just as important for the diagnosis as the lesion itself~\citep{tschandl2020human}.

Explainability is an important topic in machine learning in general and in medical imaging in particular. Saliency-based explainability methods, e.g. Grad-CAM~\citep{selvaraju2017grad}, are often used as a way to investigate if the models learn relevant features~\citep{tschandl2020human, zhang2019attention, barata2021explainable}.
Other explainability methods, such as LIME~\citep{ribeiro2016should}, Kernel-SHAP~\citep{lundberg2017unified}, and Sharp-LIME~\citep{graziani2021sharpening} are less commonly used in the medical imaging literature.

Two works, one in natural language processing and the other in dermatological imaging, have a similar approach to explainability as ours.
Within natural language processing, \cite{mathew2021hatexplain} propose a framework that explains hateful speech identification. 
Human readers were asked to identify the most important tokens in a sentence for the prediction of hateful speech.
Then, the explanation plausibility and faithfulness of the model-generated explanations were quantified by comparing to the human annotations.
Within dermatological image analysis, \cite{barata2021explainable} investigate how hierarchical taxonomies for skin lesion classification can be used to improve Conv\-Net skin cancer diagnosis capabilities.
They train networks to follow the hierarchical classification of diseases in their prediction, and to focus on relevant parts of the image.

In this work, we combine the two approaches by detecting diagnosis-explaining characteristics, each with its own localization, and train two Conv\-Nets to focus on the relevant part of the image for each characteristic.
Both networks are evaluated for the plausibility and faithfulness of their explanations.

\begin{table*}[t!]
\caption{Distribution of images over DermNetNZ and SD-260, and over the six possible diagnoses.}
\label{table:data_distribution}
\centering
\begin{tabular}{ l r r r r r r r} 
 \toprule
  & Acne & Actinic keratosis & Psoriasis & Seborrheic dermatitis & Viral warts & Vitiligo & Total\\
 \midrule
 DermNetNZ & 58 & 48 & 47 & 15 & 46 & 77 & 291\\ 
 SD-260 & 61 & 43 & 51 & 79 & 20 & 9 & 263\\
 \midrule
 Total & 119 & 91 & 98 & 94 & 66 & 86 & 554 \\
 \bottomrule
\end{tabular}
\end{table*}

\section{Material and methods}

\subsection{Explainability dataset}
To enable explainable modeling, we first identified the clinically relevant explanation taxonomy, designed an appropriate annotation protocol, and collected expert-labeled data.
This resulted in DermX\-DB: a novel dermatological explainability dataset designed to enable the training of the proposed end-to-end explainable models and quantitative explainability evaluation. 
The dataset consists of 554 images that belong to one of the following classes: acne, actinic keratosis, psoriasis, seborrheic dermatitis, viral warts, or vitiligo.
Images were sourced from DermNetNZ~\citep{dermnetnz2021} and SD-260~\citep{sun2016benchmark} with written permission from the owners.
The distribution over datasets and diseases is described in Table~\ref{table:data_distribution}.
All images were evaluated by eight board-certified dermatologists, with between four and twelve years of clinical experience. 
Each evaluation consists of a diagnosis and supporting explanations in the form of global tags, localizable characteristics, their segmentations, and additional descriptive terms for basic characteristics.



\begin{figure*}[t!]
\centering
\includegraphics[width=\textwidth]{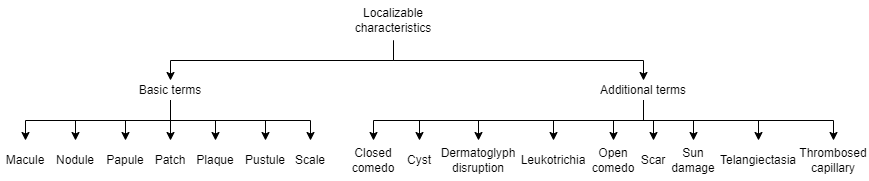}
 \caption{Localizable characteristics taxonomy.
 All characteristics were tailored to the six DermX\-DB diseases using medical resources~\citep{nast20162016, dermatology2017oakley}, and with the help of two senior dermatologists.}
\label{fig:taxonomy_localisable}
\end{figure*}

The development of this dataset included several steps. 
First, we performed several experiments to define the target diseases and the nature of the explanations.
Second, we selected the six diagnoses and defined the explanation taxonomy illustrated in Figure~\ref{fig:taxonomy_localisable}. 
Third, the labellers were allowed a short period of time to get accustomed to the annotation protocol and the labeling tool by evaluating images from an internal dataset.
Finally, DermX\-DB images were selected and sent to the dermatologists for labeling.

\paragraph{Preliminary investigation} Nine diseases were initially investigated: psoriasis, rosacea, vitiligo, seborrheic dermatitis, pityriasis rosea, viral warts, actinic keratosis, acne, and impetigo. 
These diseases were chosen based on prevalence~\citep{lim2017burden} and by the expectation that they could be diagnosed using images as the only source of patient information~\citep{dermatology2017oakley}.
Dermatologists were asked to diagnose and explain their decision in free-text for over 100 images. 
During this step, dermatologists could see the original diagnosis of the image, but had the option to disagree with it.
This step led to the exclusion of rosacea, impetigo, and pityriasis rosea from future experiments due to the difficulty in diagnosing them in the absence of the patient medical history.
It also led to the introduction of a structured ontology for the diagnosis explanations to avoid manual processing of typos and synonyms. 

\paragraph{Diagnosis and explanation ontology} Preliminary investigations also highlighted the importance of having a consistent explanation ontology. 
After analyzing free-text explanations, they were formalized as an extended list of skin lesion characteristics~\citep{nast20162016}. 
The characteristics set was selected to sufficiently explain the six target diseases~\citep{dermatology2017oakley}.
With the help of two senior dermatologists, several other relevant characteristics were added. 

The resulting set of characteristics was split into non-localizable characteristics~(e.g.~age or sex), localizable characteristics~(e.g.~plaque or open comedo), and additional descriptive terms~(e.g.~red or well-circumscribed), according to the International League of Dermatological Societies' classification~\citep{nast20162016}. 
Figure~\ref{fig:taxonomy_localisable} illustrates the final DermX\-DB explanation taxonomy, while more information about the other two types of labels is available in Appendix~Figures~\ref{fig:taxonomy_non_localisable}~and~\ref{fig:taxonomy_additional}. 

\paragraph{Annotation Protocol}
Dermatologists were first asked to diagnose the image, and then tag it with characteristics that explain their diagnosis.
No information about the gold standard diagnosis or the disease distribution was made available.
If the dermatologists were unable to evaluate the image due to poor quality, or if the image depicted a different disease than the target conditions, they had the option to discard it. 

Dermatologists could then select diagnosis-supporting non-localizable characteristics as global image tags. 
Afterwards, they could select and outline localizable characteristics. 
Dermatologists were instructed to highlight all relevant areas for each characteristic, and were only allowed to include irrelevant areas if separating them from the characteristic was too time consuming or difficult. 
In other words, they were instructed to favor sensitivity over specificity. 
Finally, basic terms~(as defined in Figure~\ref{fig:taxonomy_localisable}) could be enriched with additional descriptive terms when required for the diagnosis explanation. 
Once all tags and characteristics were added, the image could be marked as complete.

After the taxonomy and annotation protocol were defined, all dermatologists underwent two rounds of on-boarding in Darwin, a browser-based labeling tool~\citep{darwin2021}.
A screenshot of the labeling interface is shown in Appendix~Figure~\ref{fig:darwin}. 
Following this, they were asked to annotate a set of 630 images from the DermNetNZ and SD-260 datasets.

\paragraph{Data cleaning}
Once annotations were performed, the dataset went through two cleanup steps. 
First, to avoid ambiguities in the dataset, annotations with diagnoses outside the target conditions were discarded. 
This resulted in 33 images being removed from the dataset because all eight dermatologists tagged these as ‘other disease’, e.g.~acne keloidalis nuchae. 
The second step was to manually group images from the same patients. 
For all patients with more than one image, only the first image based on alphabetical order was kept. 
After cleanup, 554 images were left.
Out of all evaluations performed on these images, 150 were discarded due to reports of low image quality, resulting in 4202 individual evaluations. 

\subsection{Explainable models}
We propose two inherently explainable models for joint prediction of diagnosis and explanations.
First, we design DermX: an end-to-end clinically-inspired architecture for explainable diagnosis, and train it on the reference diagnosis and expert-identified explanation labels.
Next, we build an enhanced model that also includes learning of the explanation localization -- DermX+.
In the following, we the provide detailed description of each of the models.



\begin{figure}[ht!]
\centering
\includegraphics[width=0.65\linewidth]{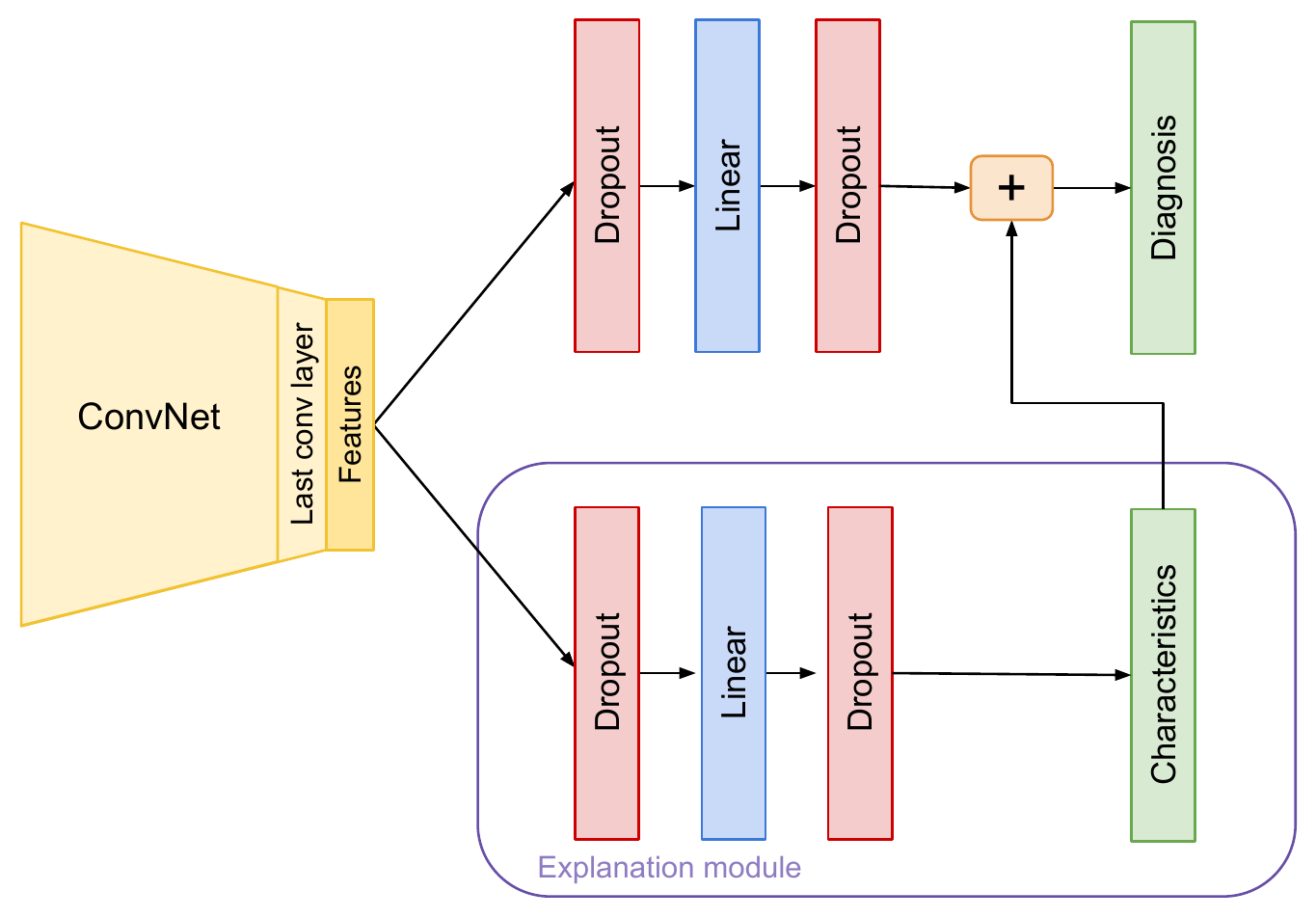}
 \caption{DermX architecture for image diagnosis with explanations in the form of skin lesion characteristics. 
 The model is trained to predict both diagnoses and characteristics.
 Image features go through a dimensionality reduction linear layer to ensure that the characteristics are not overshadowed by the image features.
 The explainability module identifies diagnosis explanations in the form of characteristics, and their localization on the image can be detected through Grad-CAMs. 
 }
\label{fig:baseline_model_architecture}
\end{figure}

\begin{figure}[ht!]
\centering
\includegraphics[width=0.65\linewidth]{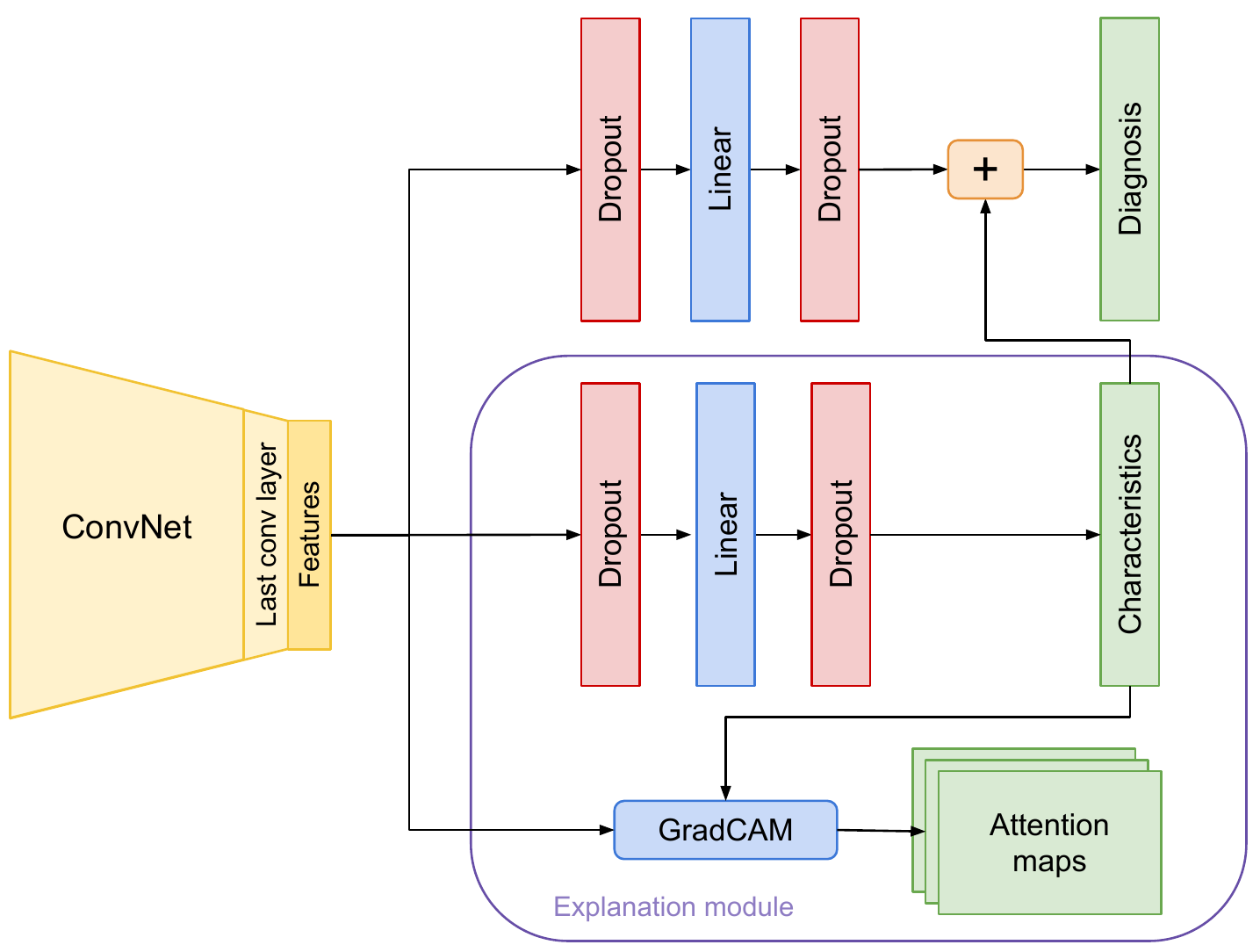}
 \caption{DermX+ architecture used to generate explanations using guided attention. 
 In addition to the DermX architecture described in Figure~\ref{fig:baseline_model_architecture}, we introduce an additional loss term for the characteristics attention map.
 The Grad-CAM attention is computed for each predicted characteristic using the features extracted by the last convolutional layer in the backbone network.
 Characteristic Grad-CAMs are then compared to the downsized fuzzy fusion masks for each characteristic.}
\label{fig:guided_attention_model_architecture}
\end{figure}


\paragraph{DermX model}
We propose a clinically-inspired model trained using the data described above.
Following the multi-task learning paradigm, the model learns how to predict a diagnosis and its supporting characteristics at the same time.
Using a Conv\-Net as an image feature extractor, we flatten and pass these features into the two prediction modules.
The explainability module passes the features through a dense block, composed of a dropout layer, a linear layer with ReLU activations, and another dropout layer.
This output is then passed into a linear layer with ten neurons and a logistic function is applied to each to give the probabilistic multi-label predictions, i.e.~multiple characteristics can be predicted at the same time.
The diagnosis module processes the image features using a similar dense block, after which they are concatenated to the characteristic logits.
For this module, the dense block also doubles as a dimensionality reduction component, allowing the image features and the characteristics to have the same order of magnitude.
The concatenated features are then passed through a linear layer with six neurons, followed by a softmax function to give our single-label prediction head for diagnoses.
Figure~\ref{fig:baseline_model_architecture} illustrates the DermX architecture.

DermX optimizes the loss defined as follows. 
Let $y_{i,d} \in \{0,1\}$ and $z_{i,c} \in \{0,1\}$ be the target diagnosis and target characteristics for image $i \in \{1,...,N\}$ in a batch of size $N$, where $d \in \{1, ..., D\}$ and $c \in \{1, ..., C\}$ denote the diagnosis and characteristic class, and let $\hat y_{i,d} \in (0,1)$ and $\hat z_{i,c} \in (0,1)$ be the diagnosis and characteristics predictions, respectively. The loss can then be written as
\begin{equation}
\label{eq:dermx_loss}
    L = \lambda_D L_D + \lambda_C L_C,
\end{equation}
where $L_D$ is the categorical cross-entropy diagnosis loss defined as 
\begin{equation}
    L_D = - \frac{1}{ND}\sum_{i=1}^{N} \sum_{d=1}^D y_{i,d} \log \hat y_{i,d},
\end{equation}
and where $L_C$ is the binary cross-entropy characteristics loss defined as
\begin{equation}
    L_C = -\frac{1}{NC}\sum_{i=1}^{N} \sum_{c=1}^C \left( z_{i,c} \log(\hat z_{i,c}) + (1 - z_{i,c}) \log{(1 - \hat z_{i,c})} \right),
\end{equation}
and where $\lambda_D$ and $\lambda_C$ are hyper-parameters for weighing the relative loss contributions.

\paragraph{DermX+ model}
We build on top of the DermX architecture by introducing a guided attention element~\citep{li2018tell}.
Figure~\ref{fig:guided_attention_model_architecture} highlights the difference between DermX and DermX+, namely the addition of a characteristic attention component.

In addition to the two losses optimized by DermX and described in Equation~\ref{eq:dermx_loss}, the DermX+ model also optimizes the attention loss term $L_A$:

\begin{equation}
    L = \lambda_D L_D + \lambda_C L_C + \lambda_A L_A,
\end{equation}
where $L_A$ is the Dice loss for attention 

\begin{equation}
   L_A = \frac{1}{NC}\sum_{i=1}^{N}\sum_{c=1}^C \left( 1 - \frac{2 A_{i,c} M_{i,c}}{A_{i,c} + M_{i,c}} \right) ,
\end{equation}
with $A_{i,c}$ being the attention map, and $M_{i,c}$ being the fuzzy localization label, both for image $i$ and characteristic $c$.


\subsection{Model training and validation}
\paragraph{Data}
Given the limited size of the dataset, we create a stratified ten-fold cross-validation setup to train explainable models, leading to approximately 500 training images and 50 test images for each fold.
Results presented in this paper are aggregated over all ten folds.
For diagnosis prediction we use the gold standard diagnosis label, as defined in the source datasets.
A characteristic was marked as relevant for a diagnosis if at least one dermatologist included the characteristic in their decision explanation.
Characteristic labels for localization were created as aggregated fuzzy maps,~i.e.~each pixel value in a mask was generated as a fraction of how many dermatologists included it in their characteristic localization.
Only characteristics selected for the correct diagnosis with regard to the gold standard were included both in defining the presence of a characteristic and in the fuzzy map aggregation. 
This way, we avoid introducing noise due to a mismatch between the diagnosis a dermatologist was explaining and the diagnosis label used to train the network.
Additionally, we exclude characteristics that appear in fewer than 30 samples throughout the dataset and characteristics with an inter-rater F1 score below 0.30.
We thus focus on closed comedo, dermatoglyph disruption, open comedo, papule, patch, plaque, pustule, scale, scar, and sun damage.

\paragraph{Implementation details}
In all experiments, we use an EfficientNet-B2~\citep{tan2019efficientnet} Conv\-Net pre-trained on the ImageNet image recognition dataset~\citep{deng2009imagenet} for feature extraction, with all layers fine-tuned on the DermX\-DB data.
Both models were trained for 93 epochs using the AdamW optimizer~\citep{loshchilov2018decoupled}, the cosine annealing with warm restarts learning rate scheduler~\citep{loshchilov2016sgdr}, and a starting learning rate of 0.0005.
Within the dense block we use linear layers with 64 neurons, dropout layers with 0.2 probability, and ReLU activations.
DermX is trained with $\lambda_D = 1$, $\lambda_C = 1$, while DermX+ uses $\lambda_D = 1$, $\lambda_C = 1$, and $\lambda_A = 10$. 
Further information about the hyper-parameters used for training and other implementation details can be found in Appendix Table~\ref{table:hyper_parameters}.

\subsection{Explainability evaluation}
We measure the performance with regard to the image diagnosis of both our dermatologists and our trained models using the F1 score, sensitivity, and specificity.
The same metrics are used to quantify the inter-rater agreement on image diagnosis and characteristics selection between dermatologists.
The model performance on characteristics is measured with regard to the fuzzy fusion label for characteristics using the same three metrics.
F1 score (also known as the Dice-Sørensen coefficient for pixel-level segmentation), sensitivity, and specificity are also used to measure the inter-rater agreement for the localizable characteristics region outlining overlap.
All values are reported as the mean and the standard deviation (std) over the 10 folds.

We define the explainability of our models as having two components: plausibility and faithfulness.
For plausibility, we focus on both the identification and the localization of characteristics.
First, we measure the F1 score, sensitivity, and specificity per characteristic to measure the models' ability to correctly identify the right explanations.
Similar to~\cite{mathew2021hatexplain}, we compare the Grad-CAM activations per characteristic with the fuzzy attention maps for each characteristic, and measure their similarity using the F1 score, sensitivity, and specificity.
All pixel-based metrics are implemented using fuzzy logic, as follows:
\begin{equation}
   \textrm{F1} = \frac{2\sum_{p \in P}{\min(A_p,M_p)}}{\sum_{p\in P}(A_p)+\sum_{p\in P}(M_p)}, 
\end{equation}

\begin{equation}
   \textrm{Sensitivity} = \frac{\sum_{p\in P}{\min(A_p,M_p)}}{\sum_{p\in P}(M_p)}, 
\end{equation}

\begin{equation}
   \textrm{Specificity} = \frac{\sum_{p\in P}{\min(1-A_p, 1-M_p)}}{\sum_{p\in P}(1-M_p)}, 
\end{equation}
where $P$ represents the pixels included in the analysis, $A$ defines the class activations, and $M$ represents the fuzzy label maps.

Following the comprehensiveness evaluation described by~\cite{deyoung2020eraser}, we measure the faithfulness of our models through the use of contrastive examples.
Given a model $m$, an input image $x$, a set of explanation outlines $e$, a contrastive image $x_e$ where all areas marked as an explanation for the image $x$ were occluded, and the class probability output $m(x)$ for the predicted class on the original input $x$ we measure the faithfulness $F$ as 
\begin{equation}
    F = m(x) - m(x_e).    
\end{equation}
In other words, the faithfulness describes what impact removing the explanations $e$ from the image would have on the decision of model $m$.
We decided not to include the sufficiency metric as it would lead to out-of-distribution images, such as a blank background with a plaque or a couple of pustules.

Finally, given the intrinsic disagreement between experts within medical fields, we postulate that explainable models should be able to properly argue their decisions, regardless of whether it matches the gold standard or not.
Similar to how dermatologists may debate the correct diagnosis for a case by highlighting different explanations that support their decision, we expect an explainable model to do the same.
However, as we do not always have the gold standard explanation for a wrong diagnosis, we need to define a basic set of explanations for any disease.
To this end, we define the expected explanation as the prevalence of each characteristic within the dermatologists explanations for a diagnosis~(Appendix~Table~\ref{table:char_disease_prevalence}).
Then, for the wrongly predicted diagnoses we compare the set of characteristics associated with that prediction with the expected explanation for the predicted diagnosis.
For example, a case incorrectly classified as psoriasis is expected to be explained using one or several of papule, plaque, and scale, which are commonly used by dermatologists in their explanations of psoriasis.
We evaluate how the model explanation for wrong diagnoses by computing the precision of the model's explanations with regard to the expected explanation for a diagnosis.

\section{Results}

\subsection{DermX\-DB analysis}

\begin{figure}[ht!]
\centering
\includegraphics[width=0.65\linewidth]{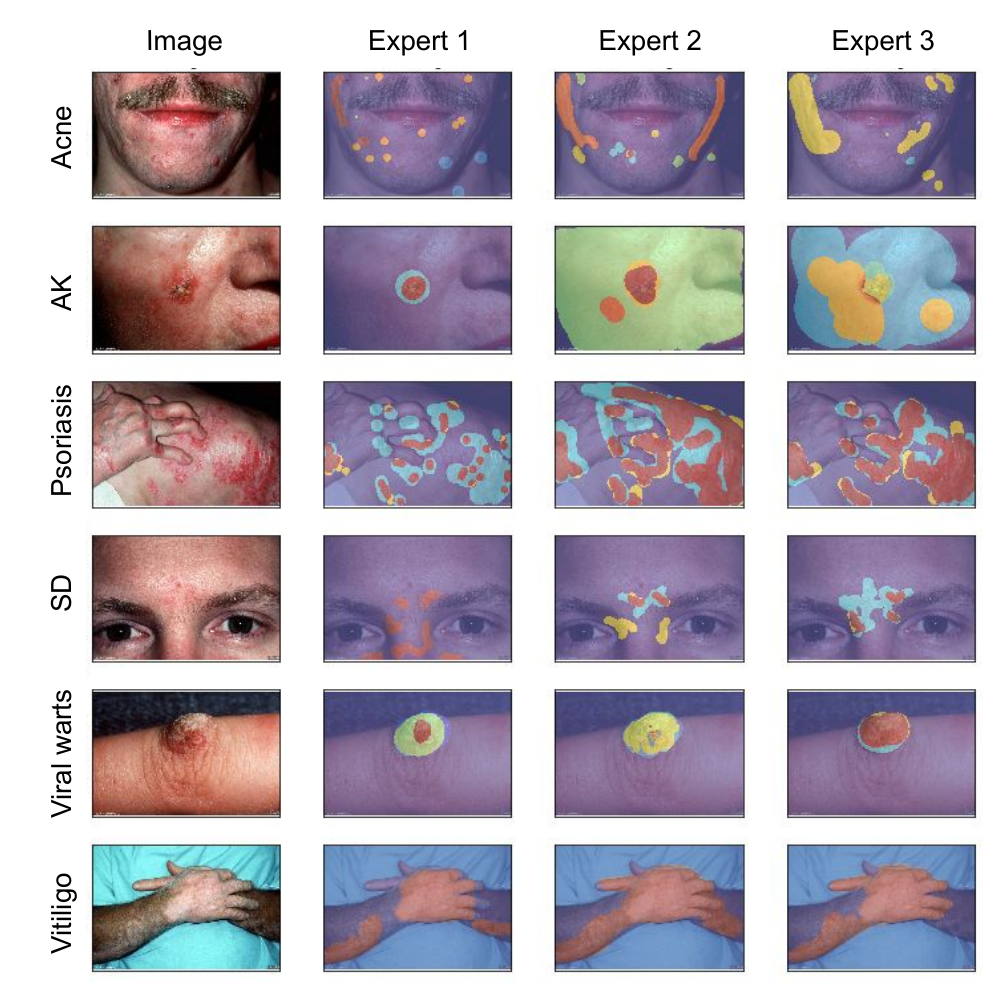} 
\caption{Differences between dermatologist-labeled attention maps, distributed over the six diseases: acne, actinic keratosis~(AK), psoriasis, seborrheic dermatitis~(SD), viral warts, and vitiligo.
The maps were computed as the union of all characteristics labeled by each of the first three dermatologists. 
Each color represents a different supporting characteristic.}
\label{fig:derm_attention}
\end{figure}

\begin{figure}[ht!]
\centering
\includegraphics[width=0.65\linewidth]{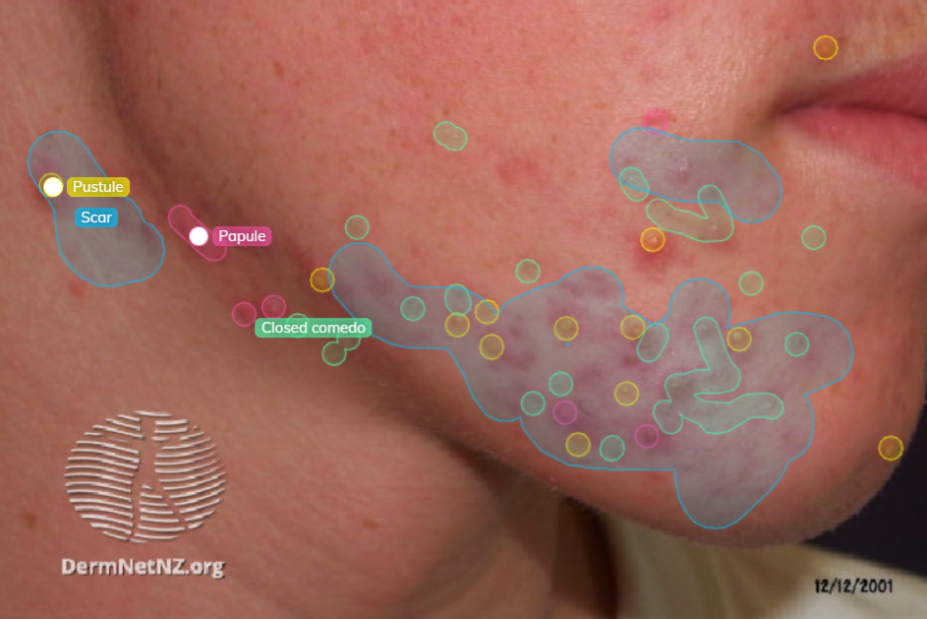} 
\caption{Characteristics labeled by a dermatologist for an acne case. Following the instructions, no characteristic was segmented but rather the region where they were present was identified without necessarily following the lesion boundaries. For more difficult characteristics to locate, e.g. scars, dermatologists were instructed to brush over entire areas containing the characteristic.}
\label{fig:chars_example}
\end{figure}


We first analyzed the data focusing on dermatologist performance with regard to the gold standard diagnosis and their inter-rater agreement on both diagnoses and supporting characteristics.
A total of 554 images were included in this analysis, each with eight evaluations performed by board-certified dermatologists. 

The dermatologist diagnostic performance in terms of mean F1 score with regard to the gold standard varies between 0.75 for seborrheic dermatitis and 0.95 for vitiligo.
Aggregated F1 scores can be seen in Table~\ref{table:comparison_diagnostic_performance_gt}.
A full description of the dermatologist performance with regard to the gold standard is available in Appendix~Table~\ref{table:derm_diagnostic_performance_gt}.


\begin{table}[t!]
\centering
\caption{Dermatologist inter-rater agreement for the presence or absence of characteristics~(mean$\pm$std). 
This analysis shows significant variation in the selection and agreement rates. 
Characteristics commonly considered important for diagnosing one of the diseases (e.g. comedones, plaques) have higher agreement rates, while uncommon characteristics (e.g. dermatoglyph disruption) display low selection and agreement rates. }

\begin{tabular}{ l r r }
 \toprule
   & F1 score & Average selection \\
 \midrule
Dermatoglyph disruption & $0.36 \pm 0.35$ & $21.50 \pm 14.34$\\
Closed comedo & $0.54 \pm 0.14$ & $41.5 \pm 22.20$\\
Open comedo  & $0.65 \pm 0.10$ & $50.88 \pm 23.28$\\ 
Papule & $0.67 \pm 0.07$ & $138.25 \pm 33.44$\\ 
Patch & $0.76 \pm 0.11$ & $114.00 \pm 43.10$\\ 
Plaque & $0.78 \pm 0.06$ & $205.12 \pm 35.98$\\ 
Pustule & $0.76 \pm 0.06$ & $58.62 \pm 13.52$\\ 
Scale  & $0.89 \pm 0.03$ & $188.50 \pm 31.16$\\
Scar & $0.46 \pm 0.14$ & $41.75 \pm 26.84$\\ 
Sun damage & $0.46 \pm 0.26$ & $31.62 \pm 14.91$\\ 
\midrule
Mean & $0.63 \pm 0.16$ & $86.42 \pm 67.05$ \\
 \bottomrule
\end{tabular}
\label{table:derm_characteristics_performance_binary}
\end{table}

Inter-rater agreement on characteristics, as described in Table~\ref{table:derm_characteristics_performance_binary}, varies significantly more, partially due to the lower number of selections per class. 
Most basic terms display high levels of agreement, with F1 scores between 0.67 and 0.89.  
The exceptions are macule with an F1 score of 0.12 and nodule with an F1 score of 0.17, both also displaying low selection rates.
Several additional terms, such as open and closed comedones, display levels of agreement similar to the basic terms.
Fig.~\ref{fig:derm_attention} illustrates an example of disagreement between three dermatologists on the location of supporting characteristics on one random case for each disease.
Additional metrics for the full set of characteristics are described in Appendix~Table~\ref{table:derm_characteristics_performance_binary_extended}.

Outlining characteristics is a more difficult task, as confirmed by the low inter-rater F1 scores reported in Table~\ref{table:derm_characteristics_performance_localisable}. 
The lower F1 values can also be explained by how difficult outlining small or poorly circumscribed characteristics is.
In terms of sensitivity, we notice the same trend as in binary agreement: dermatologists tend to agree more on the basic terms. 
Metrics for the full set of localizable characteristics are presented in Appendix~Table~\ref{table:derm_characteristics_performance_localisable_ext}.

\begin{table}[h!]
\centering
\caption{
Dermatologist inter-rater localization agreement for localizable characteristics~(mean$\pm$std).
Overlap measures show a significant variation between raters in outlining characteristics. 
Sensitivity values are high for characteristics that occupy larger areas and that often display well-circumscribed borders (e.g. plaque, scale), but tend to be lower in smaller characteristics (e.g. comedones, pustules).}

\begin{tabular}{ l r r r }
 \toprule
   & F1 score & Sensitivity & Specificity \\
 \midrule
Dermatoglyph & $0.68 \pm 0.17$ & $0.82 \pm 0.16$ & $0.98 \pm 0.03$\\ 
 disruption & & & \\
Closed comedo & $0.20 \pm 0.21$ & $0.46 \pm 0.36$ & $0.89 \pm 0.17$\\ 
Open comedo & $0.20 \pm 0.18$ & $0.44 \pm 0.33$ & $0.91 \pm 0.15$\\ 
Papule & $0.27 \pm 0.25$ & $0.49 \pm 0.34$ & $0.94 \pm 0.12$\\ 
Patch & $0.65 \pm 0.18$ & $0.80 \pm 0.19$ & $0.93 \pm 0.11$\\ 
Plaque & $0.64 \pm 0.21$ & $0.79 \pm 0.21$ & $0.93 \pm 0.10$\\ 
Pustule & $0.26 \pm 0.17$ & $0.50 \pm 0.30$ & $0.98 \pm 0.08$\\ 
Scale & $0.51 \pm 0.23$ & $0.70 \pm 0.25$ & $0.93 \pm 0.10$\\ 
Scar & $0.30 \pm 0.23$ & $0.56 \pm 0.34$ & $0.89 \pm 0.14$\\ 
Sun damage & $0.71 \pm 0.21$ & $0.84 \pm 0.19$ & $0.76 \pm 0.22$\\ 
\midrule
Mean & $0.42 \pm 0.20$ & $0.62 \pm 0.16$ & $0.91 \pm 0.06$ \\
\bottomrule

\end{tabular}
\label{table:derm_characteristics_performance_localisable}

\end{table}

\subsection{Explainable model}

\begin{table*}[ht!]
\centering
\caption{Comparison of model diagnosis performance with regard to the gold standard, presented as the mean F1 score $\pm$ std.
The models compared are the diagnosis-only model (Dx), the clinically-inspired diagnosis and characteristics model (DermX), and the DermX model trained with guided attention (DermX+).
Dermatologist scores are summarized as mean $\pm$ std across the experts.
The gold standard is the original image diagnosis as defined by the source dataset.
}

\begin{tabular}{ l r r r r } 
\toprule
   & Dx & DermX & DermX+ & Expert \\
\midrule
Acne        & $0.87 \pm 0.05$ & $0.87 \pm 0.05$ & $0.86 \pm 0.06$ & $0.94 \pm 0.02$ \\   
Actinic keratosis     & $0.80 \pm 0.06$ & $0.79 \pm 0.14$ & $0.73 \pm 0.10$ & $0.79 \pm 0.12$ \\
Psoriasis   & $0.77 \pm 0.07$ & $0.73 \pm 0.11$ & $0.80 \pm 0.09$ & $0.87 \pm 0.04$ \\   
Seborrheic dermatitis  & $0.77 \pm 0.07$ & $0.74 \pm 0.09$ & $0.74 \pm 0.10$ & $0.75 \pm 0.08$ \\   
Viral warts & $0.76 \pm 0.18$ & $0.76 \pm 0.11$ & $0.76 \pm 0.15$ & $0.92 \pm 0.05$ \\   
Vitiligo    & $0.78 \pm 0.10$ & $0.83 \pm 0.10$ & $0.86 \pm 0.08$  & $0.95 \pm 0.02$ \\    
\midrule
Mean & $0.79 \pm 0.05$ & $0.79 \pm 0.04$ & $0.79 \pm 0.04$ & $0.87 \pm 0.08$ \\
\bottomrule
\end{tabular}
\label{table:comparison_diagnostic_performance_gt}
\end{table*}

\begin{table*}[t!]
\centering
\caption{Performance comparison for characteristics identification with regard to dermatologist-generated labels, reported as mean F1 scores.
We compare the clinical diagnosis and characteristics model (DermX), the DermX model trained with guided attention (DermX+), and the inter-rater agreement among dermatologists.
A characteristic was tagged as present if at least one dermatologist marked it in an image.
The F1 score for dermatologists is based on the pairwise inter-rater agreement on characteristics (mean $\pm$ std).
}
{%
\begin{tabular}{ l r r r r}
 \toprule
   & DermX & DermX+ & Expert & Samples \\
 \midrule
Closed comedo       & $0.76 \pm 0.08$ &    $0.81 \pm 0.12$ &    $0.55 \pm 0.15$ &       96 \\
Dermatoglyph disruption &  $0.74 \pm 0.20$ &    $0.70 \pm 0.20$ &    $0.37 \pm 0.35$ &       54 \\
Open comedo   & $0.80 \pm 0.06$ &    $0.80 \pm 0.08$ &    $0.66 \pm 0.10$ &       110 \\
Papule        & $0.79 \pm 0.07$ &    $0.80 \pm 0.07$ &    $0.68 \pm 0.07$ &       278 \\
Patch         & $0.76 \pm 0.06$ &    $0.79 \pm 0.06$ &    $0.78 \pm 0.11$ &       249 \\
Plaque        & $0.88 \pm 0.03$ &    $0.90 \pm 0.03$ &    $0.79 \pm 0.06$ &       352 \\
Pustule       & $0.79 \pm 0.10$ &    $0.81 \pm 0.06$ &    $0.77 \pm 0.06$ &       106 \\
Scale         & $0.79 \pm 0.04$ &    $0.82 \pm 0.05$ &    $0.91 \pm 0.02$ &       275 \\
Scar          & $0.78 \pm 0.10$ &    $0.80 \pm 0.08$ &    $0.47 \pm 0.14$ &       115 \\
Sun damage    & $0.66 \pm 0.11$ &    $0.64 \pm 0.15$ &    $0.46 \pm 0.27$ &       78 \\
\midrule
Mean & $0.77 \pm 0.03$ & $0.79 \pm 0.03$ & $0.64 \pm 0.16$ & $171.30$ \\
 \bottomrule
\end{tabular}
}
\label{table:comparison_characteristics_performance_binary}
\end{table*}

\begin{table*}[t!]
\centering
\caption{
DermX performance for characteristics localization with regard to the fuzzy dermatologist localization maps, reported as mean soft sensitivity, specificity, and F1 score.
DermX performance metrics are computed only on samples where both the model and the dermatologists agree on the relevance of a characteristic, in order to decouple localization performance from the identification performance. 
}

\begin{tabular}{ l r r r r}
 \toprule
   & Sensitivity & Specificity & F1 score & Samples\\
 \midrule
Closed comedo       & $0.69 \pm 0.09$ & $0.69 \pm 0.05$ & $0.40 \pm 0.04$   & 75    \\
Dermatoglyph disruption & $0.69 \pm 0.09$ & $0.69 \pm 0.05$ & $0.28 \pm 0.06$   & 36    \\
Open comedo   & $0.69 \pm 0.06$ & $0.68 \pm 0.04$ & $0.36 \pm 0.05$   & 90    \\
Papule        & $0.63 \pm 0.08$ & $0.72 \pm 0.05$ & $0.34 \pm 0.04$   & 219   \\
Patch         & $0.57 \pm 0.07$ & $0.78 \pm 0.04$ & $0.43 \pm 0.05$   & 188   \\
Plaque        & $0.65 \pm 0.06$ & $0.75 \pm 0.04$ & $0.43 \pm 0.03$   & 314   \\
Pustule       & $0.69 \pm 0.07$ & $0.69 \pm 0.05$ & $0.24 \pm 0.05$   & 88    \\
Scale         & $0.65 \pm 0.07$ & $0.76 \pm 0.04$ & $0.41 \pm 0.03$   & 222   \\
Scar          & $0.64 \pm 0.06$ & $0.72 \pm 0.05$ & $0.46 \pm 0.05$   & 90    \\
Sun damage    & $0.44 \pm 0.08$ & $0.87 \pm 0.04$ & $0.56 \pm 0.06$   & 50    \\
\midrule
Mean & $0.64 \pm 0.02$ & $0.74 \pm 0.01$ & $0.39 \pm 0.02$ & $ 137.20$ \\
\bottomrule

\end{tabular}
\label{table:baseline_characteristics_performance_localisable}
\end{table*}

\begin{table*}[t!]
\centering
\caption{
DermX+ performance for characteristics localization with regard to the fuzzy dermatologist localization maps, reported as mean soft sensitivity, specificity, and F1 score.
DermX+ values are computed only on samples where both the model and the dermatologists agree on the relevance of a characteristic, in order to decouple localization performance from the identification performance. 
}

\begin{tabular}{ l r r r r}
 \toprule
   & Sensitivity & Specificity & F1 score & Samples\\
 \midrule
Closed comedo           & 	$0.11 \pm 0.08$ &	$0.96 \pm 0.02$ & $0.10 \pm 0.09$   & 63  \\
Dermatoglyph disruption & 	$0.60 \pm 0.15$ &	$0.97 \pm 0.02$ & $0.60 \pm 0.12$   & 34  \\
Open comedo             & 	$0.06 \pm 0.05$ &	$0.97 \pm 0.02$ & $0.06 \pm 0.05$   & 93  \\
Papule                  & 	$0.19 \pm 0.06$ &	$0.97 \pm 0.02$ & $0.18 \pm 0.04$   & 232 \\
Patch                   & 	$0.56 \pm 0.06$ &	$0.89 \pm 0.03$ & $0.53 \pm 0.05$   & 191 \\
Plaque                  & 	$0.65 \pm 0.06$ &	$0.91 \pm 0.01$ & $0.61 \pm 0.05$   & 312 \\
Pustule                 & 	$0.04 \pm 0.05$ &	$0.98 \pm 0.01$ & $0.03 \pm 0.03$   & 91  \\
Scale                   & 	$0.58 \pm 0.11$ &	$0.89 \pm 0.03$ & $0.49 \pm 0.07$   & 224 \\
Scar                    & 	$0.35 \pm 0.15$ &	$0.92 \pm 0.04$ & $0.35 \pm 0.12$   & 93  \\
Sun damage              & 	$0.53 \pm 0.19$ &	$0.90 \pm 0.10$ & $0.58 \pm 0.20$   & 47  \\
\midrule
Mean & $0.37 \pm 0.01$ & $0.94 \pm 0.00$ & $0.36 \pm 0.01$ & $138.00$ \\
\bottomrule

\end{tabular}
\label{table:guided_attention_characteristics_performance_localisable}
\end{table*}

We trained a clinically-inspired model from Figure~\ref{fig:dermx_model_architecture} (DermX), and the same model architecture trained with guided attention (DermX+) for characteristics localization.
We also train a diagnosis-only model (Dx) to check whether adding explanations impact the diagnosis performance of DermX and DermX+.
Table~\ref{table:comparison_diagnostic_performance_gt} compares the diagnostic performance between all three models and the dermatologists with regard to the gold standard diagnosis.
More information about their diagnostic performance is presented in~\ref{appendix:model}.
For comparison, we trained a diagnosis-only model with a ResNet50~\citep{he2016deep} base to validate the choice of architecture, and a diagnosis-only model trained with proportional class weights.
The ResNet-based model achieved a macro F1-score of $0.79 \pm 0.06$, while the weighted class model showed a similar macro F1-score of $0.78 \pm 0.05$.
More information about these two models is available in Appendix~Table~\ref{table:comparison_diagnostic_performance_gt_ext}.
Additionally, we trained four interpretable models on the characteristics data for diagnosis prediction: a logistic regression model, a decision tree, a k-nearest neighbor with five neighbors, and a categorical naive Bayes models.
These models obtained macro F1 scores of $0.86 \pm 0.04$, $0.85 \pm 0.05$, $0.80 \pm 0.05$, and $0.86 \pm 0.05$, respectively.

All models display similar F1 scores on all six diseases.
The best results are obtained for vitiligo and acne, two disease classes where dermatologists also display high F1 score values.
Seborrheic dermatitis on the other hand seems to be a difficult disease class for both dermatologists and models.
For the rest of the results section we will focus on DermX and DermX+.

In terms of explanation plausibility, we look at both the identification of explanations, defined as the ability to detect the same characteristics as a dermatologist, and at their localization in the image.
A comparison of F1 scores is described in Table~\ref{table:comparison_characteristics_performance_binary}. 
The two models perform well for explanation identification, with DermX+ obtaining slightly better results on most characteristics.
Compared to dermatologists, the models perform within standard deviation bounds of the inter-rater agreement.
Additional metrics are reported in Appendix~Tables~\ref{table:baseline_characteristics_performance_binary}~and~\ref{table:guided_attention_characteristics_performance_binary}.

\begin{figure}[t!]
\centering
\includegraphics[width=0.65\linewidth]{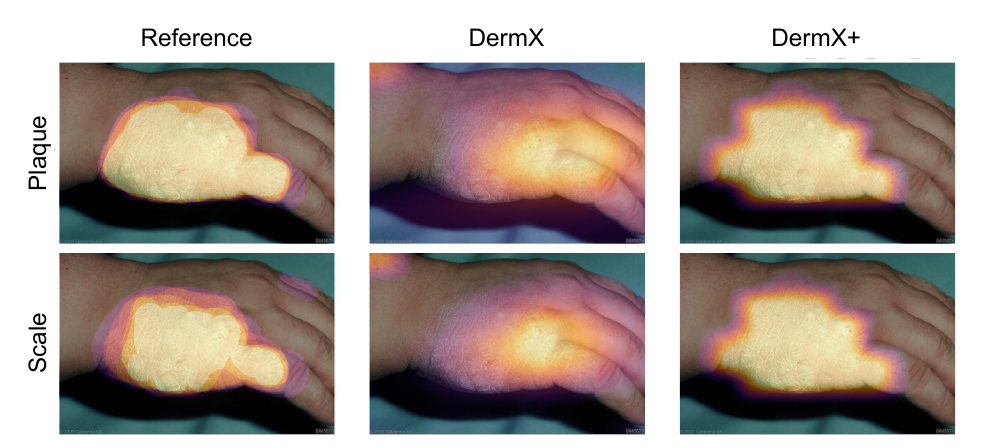}
\caption{Characteristic attention maps for a correctly classified psoriasis case, for the two identified characteristics: plaque (first row) and scale (second row).
The first column shows the dermatologist-derived fuzzy attention map, the second one illustrates the Grad-CAM for each characteristic generated by DermX, while the last column shows the DermX+ Grad-CAM maps.
DermX+ displays much closer results to the gold standard, while DermX maps include more irrelevant information, such as finger knuckles.
}
\label{fig:model_attention_example}
\end{figure}

\begin{figure}[t!]
\centering
\includegraphics[width=0.65\linewidth]{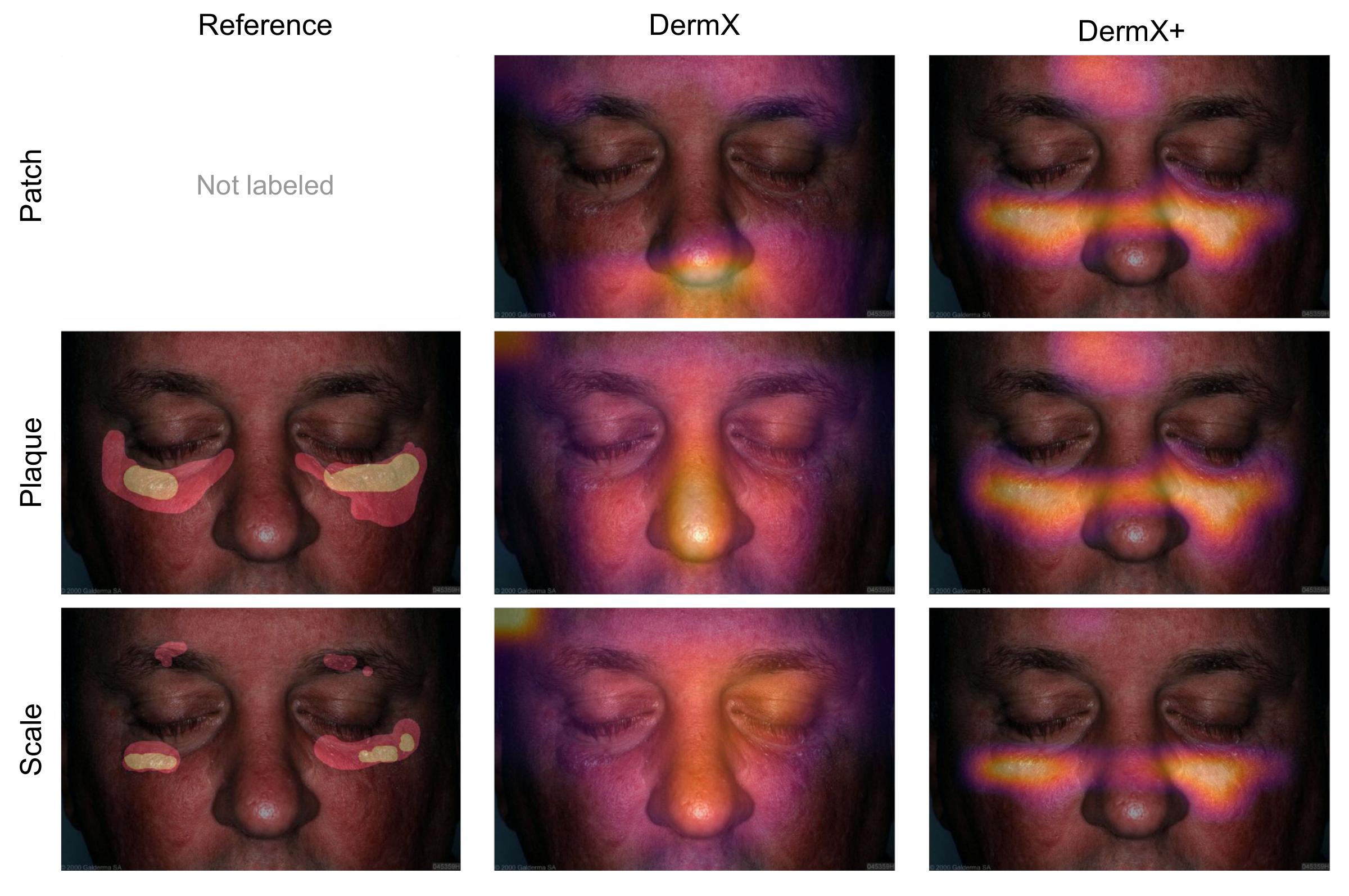}
\caption{Characteristic attention maps for a psoriasis case wrongly classified as seborrheic dermatitis by both DermX and DermX+.
While both models detect plaque and scale in the image, their detection of patch, a characteristic indicative of seborrheic dermatitis, leads them to misdiagnose the image. 
}
\label{fig:model_attention_wrong_pred}
\end{figure}

The localization plausibility of the models' explanation is quantified in Tables~\ref{table:baseline_characteristics_performance_localisable}~and~\ref{table:guided_attention_characteristics_performance_localisable}, with more statistics being presented in Appendix~Tables~\ref{table:baseline_characteristics_performance_localisable_extra},~\ref{table:baseline_characteristics_performance_localisable_extra_wrong},~\ref{table:guided_attention_characteristics_performance_localisable_extra},~and~\ref{table:guided_attention_characteristics_performance_localisable_extra_wrong}.
DermX performs adequately well on all characteristics.
DermX+ is better at localizing large characteristics,~e.g. patches or scales, but performs poorly on smaller characteristics, e.g. open and closed comedones.
Dermatologists F1 scores indicate that the two models are, in some characteristics, within standard deviation of the inter-rater agreement.
For other characteristics, such as dermatoglyph disruption for DermX and pustule for DermX+, the model performance is below the expert inter-rater agreement.
Figure~\ref{fig:model_attention_example} illustrates the explanations given for a correctly predicted psoriasis case by DermX and DermX+, respectively, while Figure~\ref{fig:model_attention_wrong_pred} shows the explanations given by the two ConvNets for a misclassified psoriasis case.

Explanation precision scores for the correct diagnosis prediction were computed with regard to the dermatologist labels. 
The resulting values are $0.88 \pm 0.03$ for DermX and $0.90 \pm 0.03$ for DermX+.
On the wrong diagnosis prediction, DermX precision is $0.85 \pm 0.06$, while DermX+ precision is $0.86 \pm 0.04$.
Mean faithfulness results are $0.42 \pm 0.06$ for DermX and $0.27 \pm 0.06$ for DermX+.

\section{Discussion}
To the best of our knowledge, DermX is the first end-to-end framework created for the purpose of explaining automated dermatological diagnoses.
The two Conv\-Nets we introduce, DermX and DermX+, mimic the dermatological approach to diagnosing skin conditions: first they recognize supporting characteristics, then they use these characteristics as well as other high level information to arrive at a diagnosis.
In addition to identifying supporting characteristics as explanations to a diagnosis, DermX+ also learns the localization of the explanations via the guided attention loss.
The decision to use an attention mechanism for localization rather than a semantic segmentation approach was guided by the design of the annotation protocol. 
Because dermatologists were instructed to highlight explanation regions in an image with a focus on sensitivity instead of specificity, the resulting outlines are not well suited as segmentation masks.
For this work to be possible, we collected diagnoses and supporting characteristics for 554 images from eight board-certified dermatologists.

During the process of collecting the DermX\-DB data, we found that dermatologists often focus on different characteristics when diagnosing a case.
While most explanations for diseases display a set of common characteristics, such as scales, plaques, and papules for psoriasis, there is also a long tail of relevant characteristics that are not always selected.
In addition, we found that inter-rater agreement was low for characteristics localization.
This may be caused by the difficulty in outlining characteristics with poorly defined boundaries, such as patch, but also by dermatologists differing in their approach to outlining smaller characteristics, such as open and closed comedones.

The contrast between high agreement on diagnoses and low agreement on supporting characteristics illustrates how different experts perceive explanations in different ways. 
Although they generally agree on the diagnosis, dermatologists focus on different characteristics to explain their decision. 
To properly evaluate a model's explanations, we must therefore consider the opinions of multiple experts.
Moreover, this intrinsic variability in how experts approach explanations lends more urgency to the need for quantifiable explanation methods.

From a modeling perspective, our results contradict the common adage that there must be a trade-off between predictive power and explainability.
DermX and DermX+ both report the same diagnosis performance as a standard diagnosis-only Conv\-Net, while also offering plausible explanations for their decisions.
Even in cases where they predict the wrong diagnosis, both models provide arguments that make sense for their prediction.
Most explanations given by both models are within standard deviation of the inter-rater agreement on characteristics, suggesting that either model may function as a second opinion with realistic decision explanations.

When compared to interpretable models trained on the characteristics data, both DermX and DermX+ obtain a diagnosis performance within standard deviation of the models using manually labeled features.
None of the models we trained obtains a diagnosis performance as high as that of experts.
We postulate that this is due to the difficulty of the dataset, as shown by the inter-rater agreement in Table~\ref{table:comparison_diagnostic_performance_gt}, and due to the limited amount of training data.
On the other hand, our results are on par with the diagnosis accuracy reported by other research groups using dermatological clinical photography, which varies between 56.7\% on 134 classes~\citep{han2020augmented} and 86.53\% on four classes~\citep{burlina2019automated}.

Our localization results for both models are lower than the inter-rater agreement on expert-derived maps for most characteristics.
This may in part be due to the low inter-rater agreement on the localization data, and in part due to the small scale at which the maps were computed (nine by nine pixels for the EfficientNet-B2 architecture).
However, the high sensitivity values show that these maps are often good enough to give a visual hint as to the location of the characteristic in an image.
Such a hint would be useful in cases where an expert using DermX or DermX+ as a second opinion did not notice that characteristic.
Comparing the two models, DermX+ displays lower overall F1 scores than DermX, while showcasing higher overall specificity and high sensitivity on large characteristics.
This may be explained by its training target: dermatologist attention maps were linearly scaled down to the size of the feature maps, which may have reduced the target attention map of small characteristics to an almost empty mask.
Another possible explanation is DermX's reliance on the sometimes noisy localization data.
In particular, for characteristics smaller than 1cm (closed comedo, open comedo, papule, and pustule), DermX+ is clearly outperformed by DermX due to the lower specificity and higher sensitivity of DermX.
In the future, we plan on investigating different ways of downscaling the masks, and to increase the feature map size to take advantage of the high resolution gold standard attention maps.

\begin{figure}[t!]
\centering
\includegraphics[width=0.65\linewidth]{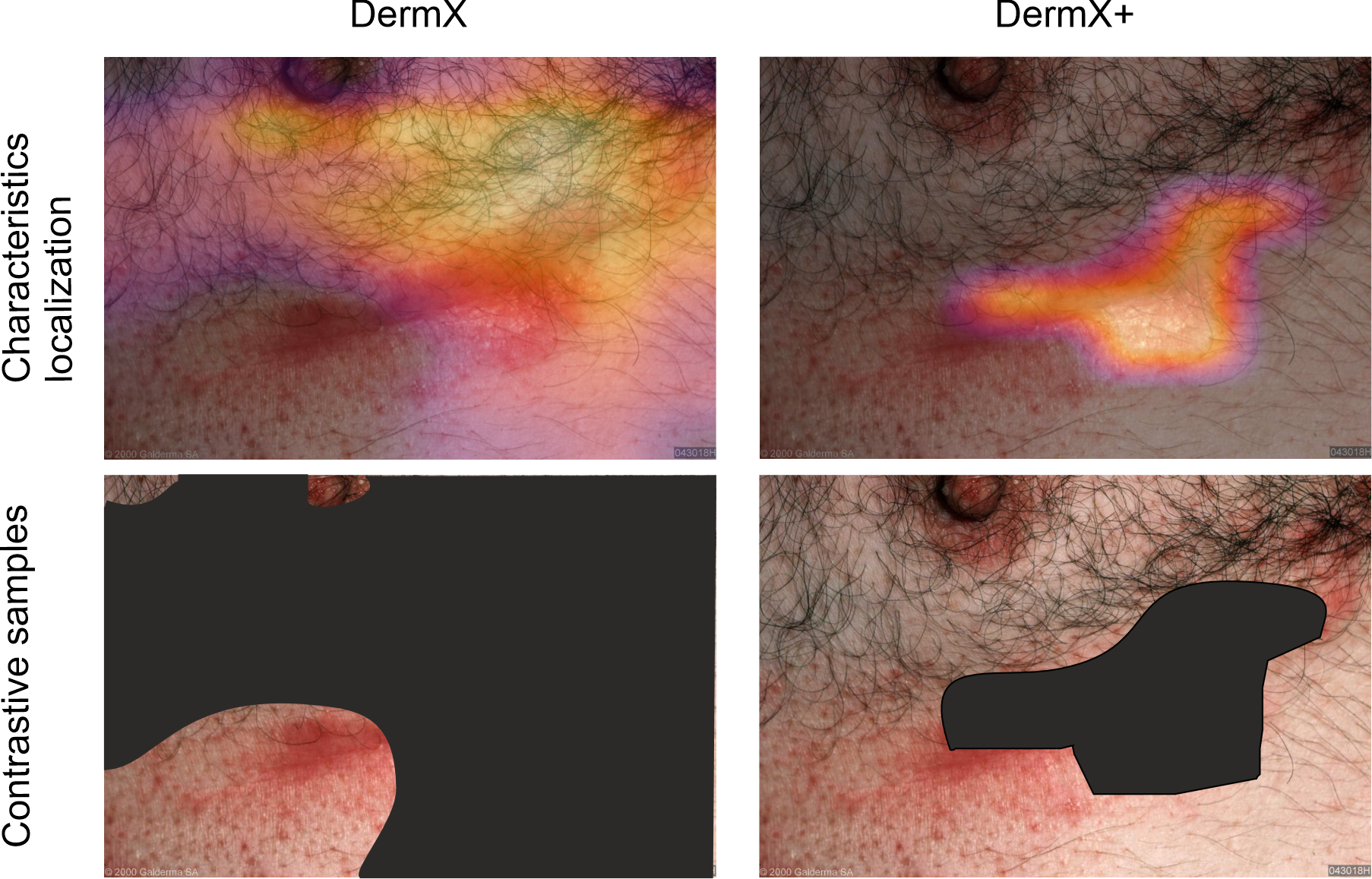}
\caption{Contrastive samples on a DermXDB psoriasis image for DermX and DermX+.
DermX, a more sensitive and less specific model, occludes large parts of the image, while
DermX+, a more specific and less sensitive model, occludes only the lesion.
When evaluating the contrastive sample for DermX+, the model has the possibility to use other diagnosis hints in the image (defined in DermXDB as non-localizable characteristics) that are occluded in the DermX sample.
}
\label{fig:faithfulness}
\end{figure}

Mean faithfulness scores above zero for both models prove that the characteristic localizations are indeed explanations about the diagnosis decision mechanisms of the models.
DermX+, a more specific model in terms of characteristic localization, has lower faithfulness scores than DermX, which tends to include adjacent regions in its localizations.
Figure~\ref{fig:faithfulness} showcases the impact a model's specificity and sensitivity have on the contrastive samples, and therefore on the faithfulness metric.
In this example, the contrastive sample created by DermX+ still displays image-level non-localizable characteristics, information which is occluded in the DermX contrastive sample.
This further confirms the importance of image-level tags in skin lesion diagnosis, e.g. by noticing that acne is predominantly located on the face or upper trunk, or that actinic keratosis most commonly affects elderly people.

This work opens many new research avenues in the domain of medical image diagnosis explainability.
From a dermatological data perspective, we plan on adding more diseases and supporting characteristics to DermX\-DB.
The annotation protocol developed as part of DermX\-DB can serve as an inspiration not only for explaining other dermatological diseases, but also for different radiology and pathology investigations.
In radiology, imaging findings are routinely recorded, including the supporting characteristics for the diagnosis. 
For example, the malignancy of a lesion seen on a mammogram could be supported by localized characteristics such as calcifications and dense tissue~\citep{birads_acr}. 
The network architectures we proposed could also be applied to learning the supporting radiological findings as explanations to diagnoses provided appropriately labeled datasets. 
From a modeling perspective, we will focus on leveraging the full potential of DermX\-DB by adding image-level explanations to the diagnosis models, and by incorporating the additional descriptive terms into the explanation setup.
More work can be done in improving the characteristic localization.
We will be focusing in particular on introducing the adversarial loss described in~\cite{li2018tell} for semi-supervised attention guidance.
Another approach we will be to train object detection networks~\citep{tan2020efficientdet, redmon2018yolov3} to detect the supporting characteristics alongside the diagnosis.
Once the localization reaches a higher performance, a true test of the DermX architecture would be to set up a clinical trial where its predictions would be used as a second opinion for health care professionals of various levels of expertise.

\section{Conclusions}
In this work, we introduce DermX -- a novel, clinically-inspired explainable Conv\-Net architecture for skin lesion diagnosis.
We also introduce a variation named DermX+ that adds a guided attention loss such that localization of lesion characteristics becomes a part of the supervised training.
We quantify the explanation quality by comparing it with explanations given by board-certified dermatologists with different levels of clinical experience.
To facilitate future work, we release this explainability dataset to the public, and describe the annotation protocol used for its creation.

\section*{Acknowledgments} 
Raluca Jalaboi's work was supported in part by the Danish Innovation Fund under Grant 0153-00154A. 
Ole Winther’s work was funded in part by the Novo Nordisk Foundation through the Center for Basic Machine Learning Research in Life Science (NNF20OC0062606).
Ole Winther acknowledges supporting the Pioneer Centre for AI, DNRF grant number P1.

\typeout{}

\bibliographystyle{unsrtnat}
\bibliography{main}

\clearpage
\onecolumn

\appendix

\section{Additional dataset information}
\label{appendix:dataset}

\begin{figure}[h!]
\centering
\includegraphics[width=1\textwidth]{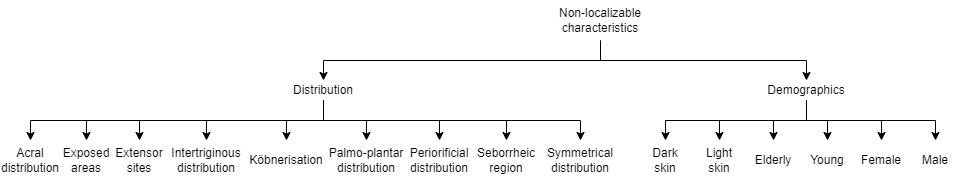} 
 \caption{
 Non-localizable characteristics taxonomy. These characteristics were added to the International League of Dermatological Societies' classification as global image tags after being flagged as relevant by our senior dermatologists.}
\label{fig:taxonomy_non_localisable}
\end{figure}

\begin{figure}[h!]
\centering
\includegraphics[width=\textwidth]{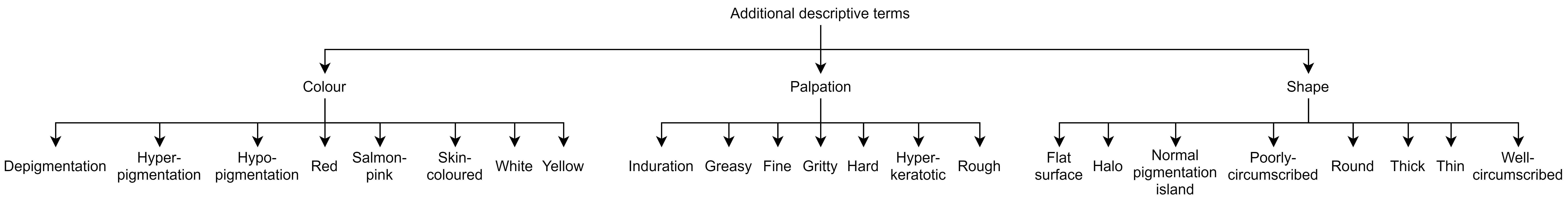} 

 \caption{Additional descriptive terms for localizable characteristics. All terms were tailored for the six diseases from medical resources~\citep{nast20162016, dermatology2017oakley}, and with the help of two senior dermatologists.}
\label{fig:taxonomy_additional}
\end{figure}

\begin{figure}[h!]
\centering
\includegraphics[width=\textwidth]{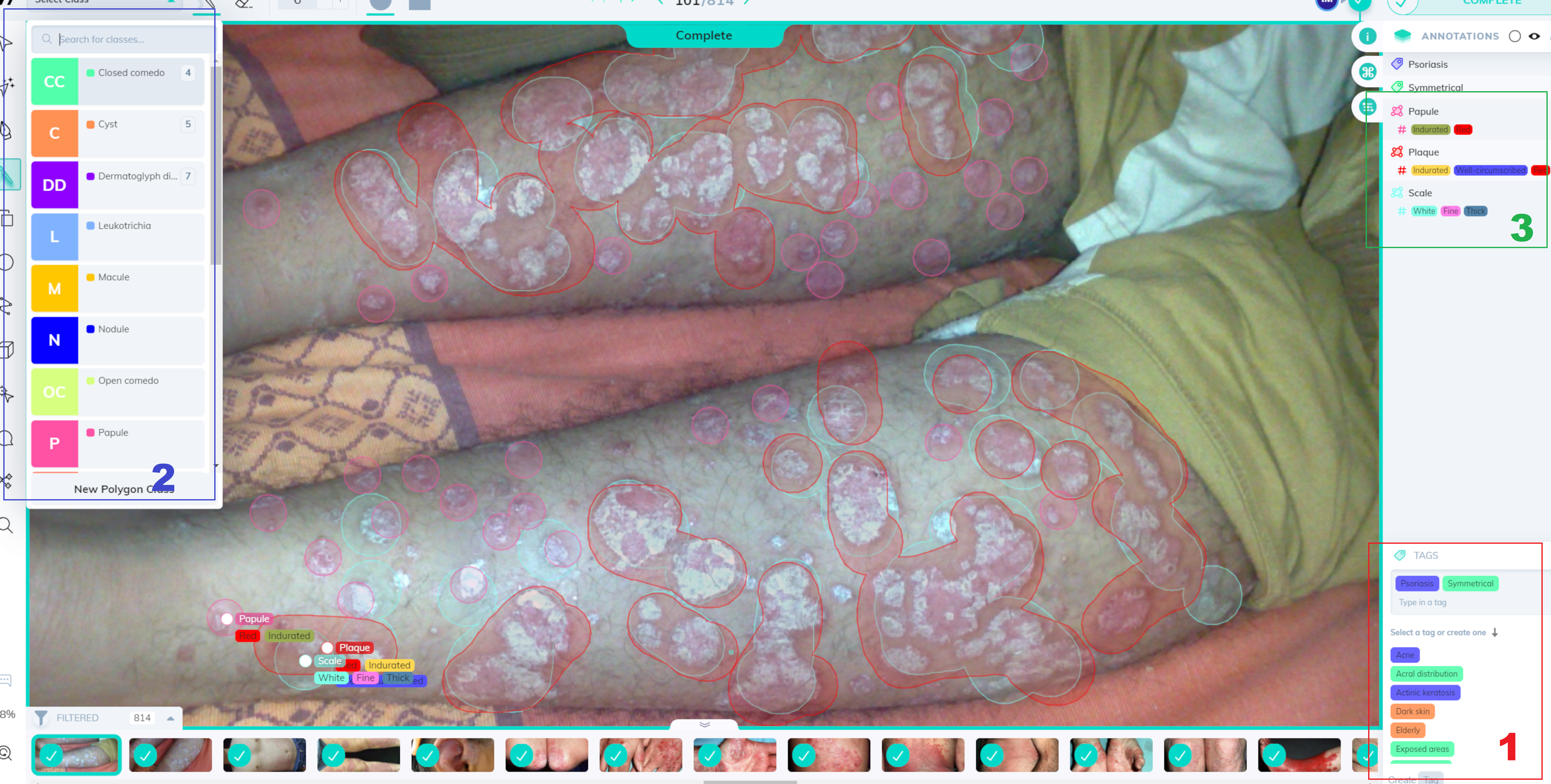} 
\caption{Labeling tool interface, exemplified for a psoriasis case from the SD-260 dataset. 
In the global tag search box (area 1, bottom right), dermatologists can select the disease, relevant demographics information, and lesion distribution. 
The brush selection menu (area 2, top left) allows them to select and mark localizable characteristics on the image.
The full annotation menu (area 3, top right) is used to select of additional descriptive terms for the localized basic terms.}
\label{fig:darwin}
\end{figure}

\begin{table}[h!]
\centering
\caption{Dermatologist diagnosis performance with regard to the gold standard~(mean$\pm$std). }

\begin{tabular}{ l r r r r } 
\toprule
  & F1 score & Sensitivity & Specificity & Average selection \\
\midrule
Acne & $0.94 \pm 0.02$ & $0.90 \pm 0.04$ & $0.99 \pm 0.00$ & $102.75 \pm 10.00$\\ 
Actinic keratosis & $0.79 \pm 0.12$ & $0.68 \pm 0.16$ & $1.00 \pm 0.00$ & $57.62 \pm 16.54$\\ 
Psoriasis & $0.87 \pm 0.04$ & $0.88 \pm 0.04$ & $0.97 \pm 0.02$ & $98.00 \pm 8.09$\\ 
Seborrheic dermatitis & $0.75 \pm 0.08$ & $0.64 \pm 0.11$ & $0.99 \pm 0.01$ & $60.38 \pm 9.23$\\ 
Viral warts & $0.92 \pm 0.05$ & $0.85 \pm 0.08$ & $1.00 \pm 0.00$ & $55.38 \pm 5.05$\\ 
Vitiligo & $0.95 \pm 0.02$ & $0.90 \pm 0.05$ & $1.00 \pm 0.00$ & $76.25 \pm 4.82$ \\
\midrule
Mean & $0.87 \pm 0.08$ & $0.81 \pm 0.11$ & $0.99 \pm 0.01$ & $75.06 \pm 19.15$ \\
\bottomrule
\end{tabular}

\label{table:derm_diagnostic_performance_gt}
\end{table}

\begin{table}[h!]
\centering
\caption{Dermatologist inter-rater agreement for the presence or absence of characteristics~(mean$\pm$std). This analysis shows significant variation in the selection and agreement rates. 
Characteristics commonly considered important for diagnosing one of the diseases (e.g. comedones, plaques) have higher agreement rates, while uncommon characteristics (e.g. leukotrichia, telangiectasia) display low selection and agreement rates. }

\begin{tabular}{ l r r r r r }
 \toprule
   & F1 score & Sensitivity & Specificity & Cohen's kappa & Average selection \\
 \midrule
 \textbf{Basic terms}\\
Macule & $0.12 \pm 0.15$ & $0.18 \pm 0.23$ & $0.95 \pm 0.07$ & $0.09 \pm 0.15$ & $25.25 \pm 30.37$\\ 
Nodule & $0.17 \pm 0.16$ & $0.26 \pm 0.27$ & $0.97 \pm 0.04$ & $0.16 \pm 0.15$ & $17.88 \pm 17.42$\\ 
Papule & $0.67 \pm 0.07$ & $0.69 \pm 0.13$ & $0.86 \pm 0.08$ & $0.52 \pm 0.09$ & $138.25 \pm 33.44$\\ 
Patch & $0.76 \pm 0.11$ & $0.79 \pm 0.18$ & $0.92 \pm 0.08$ & $0.68 \pm 0.15$ & $114.00 \pm 43.10$\\ 
Plaque & $0.78 \pm 0.06$ & $0.80 \pm 0.11$ & $0.84 \pm 0.08$ & $0.62 \pm 0.09$ & $205.12 \pm 35.98$\\ 
Pustule & $0.76 \pm 0.06$ & $0.78 \pm 0.11$ & $0.96 \pm 0.02$ & $0.73 \pm 0.06$ & $58.62 \pm 13.52$\\ 
Scale  & $0.89 \pm 0.03$ & $0.90 \pm 0.06$ & $0.93 \pm 0.04$ & $0.82 \pm 0.04$ & $188.50 \pm 31.16$\\ 
\midrule
 \textbf{Additional terms}\\
Closed comedo & $0.54 \pm 0.14$ & $0.64 \pm 0.25$ & $0.96 \pm 0.04$ & $0.51 \pm 0.14$ & $41.5 \pm 22.20$\\ 
Cyst & $0.20 \pm 0.16$ & $0.31 \pm 0.34$ & $0.99 \pm 0.01$ & $0.19 \pm 0.16$ & $6.25 \pm 6.70$\\ 
Dermatoglyph disruption & $0.36 \pm 0.35$ & $0.39 \pm 0.38$ & $0.97 \pm 0.03$ & $0.37 \pm 0.34$ & $21.50 \pm 14.34$\\ 
Leukotrichia & $0.43 \pm 0.41$ & $0.45 \pm 0.43$ & $1.00 \pm 0.01$ & $0.45 \pm 0.41$ & $4.62 \pm 2.69$\\ 
Open comedo  & $0.65 \pm 0.10$ & $0.71 \pm 0.21$ & $0.96 \pm 0.04$ & $0.61 \pm 0.11$ & $50.88 \pm 23.28$\\ 
Scar & $0.46 \pm 0.14$ & $0.57 \pm 0.26$ & $0.95 \pm 0.05$ & $0.42 \pm 0.14$ & $41.75 \pm 26.84$\\ 
Sun damage & $0.46 \pm 0.26$ & $0.53 \pm 0.29$ & $0.97 \pm 0.02$ & $0.43 \pm 0.25$ & $31.62 \pm 14.91$\\ 
Telangiectasia & $0.17 \pm 0.25$ & $0.19 \pm 0.28$ & $0.99 \pm 0.01$ & $0.19 \pm 0.25$ & $6.12 \pm 5.60$\\ 
Thrombosed capillaries & $0.36 \pm 0.27$ & $0.45 \pm 0.37$ & $0.98 \pm 0.02$ & $0.35 \pm 0.26$ & $15.88 \pm 11.78$\\ 
\midrule
Mean & $0.49 \pm 0.24$ & $0.54 \pm 0.22$ & $0.95 \pm 0.04$ & $0.45 \pm 0.21$ & $60.48 \pm 62.97$ \\
 \bottomrule
\end{tabular}
\label{table:derm_characteristics_performance_binary_extended}

\end{table}

\begin{table}[h!]
\centering
\caption{
Dermatologist inter-rater localization agreement for localizable characteristics~(mean$\pm$std).
Overlap measures show a significant variation between raters in outlining characteristics. 
Sensitivity values are high for characteristics that occupy larger areas and that often display well-circumscribed borders (e.g. plaque, scale), but tend to be lower in smaller characteristics (e.g. comedones, pustules).}

\begin{tabular}{ l r r r }
 \toprule
   & F1 score & Sensitivity & Specificity \\
 \midrule
\textbf{Basic terms}\\
Macule & $0.21 \pm 0.16$ & $0.44 \pm 0.3$ & $0.95 \pm 0.11$\\ 
Nodule & $0.31 \pm 0.24$ & $0.55 \pm 0.33$ & $0.96 \pm 0.09$\\ 
Papule & $0.27 \pm 0.25$ & $0.49 \pm 0.34$ & $0.94 \pm 0.12$\\ 
Patch & $0.65 \pm 0.18$ & $0.80 \pm 0.19$ & $0.93 \pm 0.11$\\ 
Plaque & $0.64 \pm 0.21$ & $0.79 \pm 0.21$ & $0.93 \pm 0.10$\\ 
Pustule & $0.26 \pm 0.17$ & $0.50 \pm 0.30$ & $0.98 \pm 0.08$\\ 
Scale & $0.51 \pm 0.23$ & $0.70 \pm 0.25$ & $0.93 \pm 0.10$\\ 
\midrule
\textbf{Additional terms}\\
Closed comedo & $0.20 \pm 0.21$ & $0.46 \pm 0.36$ & $0.89 \pm 0.17$\\ 
Cyst & $0.39 \pm 0.27$ & $0.59 \pm 0.31$ & $0.98 \pm 0.07$\\ 
Dermatoglyph disruption & $0.68 \pm 0.17$ & $0.82 \pm 0.16$ & $0.98 \pm 0.03$\\ 
Leukotrichia & $0.50 \pm 0.14$ & $0.70 \pm 0.21$ & $0.96 \pm 0.07$\\ 
Open comedo & $0.20 \pm 0.18$ & $0.44 \pm 0.33$ & $0.91 \pm 0.15$\\ 
Scar & $0.30 \pm 0.23$ & $0.56 \pm 0.34$ & $0.89 \pm 0.14$\\ 
Sun damage & $0.71 \pm 0.21$ & $0.84 \pm 0.19$ & $0.76 \pm 0.22$\\ 
Telangiectasia & $0.29 \pm 0.14$ & $0.53 \pm 0.28$ & $0.93 \pm 0.12$ \\
Thrombosed capillaries & $0.42 \pm 0.22$ & $0.65 \pm 0.28$ & $0.99 \pm 0.02$\\
\midrule
Mean & $0.41 \pm 0.18$ & $0.62 \pm 0.14$ & $0.93 \pm 0.05$ \\
\bottomrule

\end{tabular}
\label{table:derm_characteristics_performance_localisable_ext}

\end{table}

\begin{table}[h!]
\centering
\caption{Characteristics prevalence per disease.}

\begin{tabular}{ l r r r r r r }
 \toprule
 & Acne & Actinic keratosis & Psoriasis & Seborrheic dermatitis & Viral warts & Vitiligo \\
 \midrule

Closed comedo  & 0.40 & 0.00 & 0.00 & 0.00 & 0.00 & 0.00 \\
Dermatoglyph disruption & 0.00 & 0.00 & 0.00 & 0.00 & 0.38 & 0.00 \\
Open comedo & 0.49 & 0.00 & 0.00 & 0.01 & 0.00 & 0.00 \\
Papule & 0.70 & 0.13 & 0.18 & 0.04 & 0.65 & 0.00 \\
Patch  & 0.02 & 0.32 & 0.02 & 0.24 & 0.00 & 0.97 \\
Plaque & 0.04 & 0.59 & 0.96 & 0.71 & 0.46 & 0.01 \\
Pustule & 0.56 & 0.00 & 0.01 & 0.00 & 0.00 & 0.00 \\
Scale &  0.01 & 0.81 & 0.89 & 0.78 & 0.05 & 0.00 \\
Scar & 0.38 & 0.03 & 0.00 & 0.00 & 0.00 & 0.00 \\
Sun damage & 0.00 & 0.47 & 0.01 & 0.01 & 0.00 & 0.01 \\
\bottomrule

\end{tabular}
\label{table:char_disease_prevalence}

\end{table}

\section{Model training and extended performance}
\label{appendix:model}
\begin{table}[h!]
    \centering
    \caption{Training hyper-parameters common to all trained models.}
    \label{table:hyper_parameters}
    \begin{tabular}{lrr}
    \toprule
    Hyperparameter & Value \\
    \midrule
        Batch size & 32 \\
        Rotation & 10 \\
        Zoom & 0.15 \\
        Brightness & 0.35 \\
        Contrast & 0.20 \\
        Saturation & 0.20 \\
        Scale & (0.85, 1.15) \\
        Translate & (0.15, 0.15) \\
        Hue & 0.15 \\
        Dropout & 0.2 \\
    \bottomrule
    \end{tabular}
\end{table}

\begin{table*}[h!]
    \centering
    \caption{Comparison of model diagnosis performance with regard to the gold standard, presented as the mean F1 score ± std. The models compared
are the diagnosis-only model (Dx), diagnosis-only with a ResNet50 base (DxRN), a diagnosis-only model trained with class weights, the clinically-inspired
diagnosis and characteristics model (DermX), and the DermX model trained with guided attention (DermX+). Dermatologist scores are summarized as
mean ± std across the experts. The gold standard is the original image diagnosis as defined by the source dataset.}
\begin{tabular}{ l r r r r r r } 
\toprule
   & Dx & DxRN & DxW & DermX & DermX+ & Expert \\
\midrule
Acne        & $0.87 \pm 0.05$ & $0.90 \pm 0.06$ & $0.85 \pm 0.05$ & $0.87 \pm 0.05$ & $0.86 \pm 0.06$ & $0.94 \pm 0.02$ \\   
Actinic keratosis     & $0.80 \pm 0.06$ & $0.74 \pm 0.16$ & $0.77 \pm 0.15$ & $0.79 \pm 0.14$ & $0.73 \pm 0.10$ & $0.79 \pm 0.12$ \\
Psoriasis   & $0.77 \pm 0.07$ & $0.76 \pm 0.07$ & $0.77 \pm 0.06$ & $0.73 \pm 0.11$ & $0.80 \pm 0.09$ & $0.87 \pm 0.04$ \\   
Seborrheic dermatitis  & $0.77 \pm 0.07$ & $0.72 \pm 0.11$ & $0.75 \pm 0.14$ & $0.74 \pm 0.09$ & $0.74 \pm 0.10$ & $0.75 \pm 0.08$ \\   
Viral warts & $0.76 \pm 0.18$ & $0.74 \pm 0.16$ & $0.73 \pm 0.20$ & $0.76 \pm 0.11$ & $0.76 \pm 0.15$ & $0.92 \pm 0.05$ \\   
Vitiligo    & $0.78 \pm 0.10$ & $0.85 \pm 0.06$  & $0.82 \pm 0.07$ & $0.83 \pm 0.10$ & $0.86 \pm 0.08
$  & $0.95 \pm 0.02$ \\    
\midrule
Mean & $0.79 \pm 0.05$ & $0.79 \pm 0.06$ & $0.78 \pm 0.05$ & $0.79 \pm 0.04$ & $0.79 \pm 0.04$ & $0.87 \pm 0.08$ \\
\bottomrule
\end{tabular}
\label{table:comparison_diagnostic_performance_gt_ext}
\end{table*}

\begin{table}[h!]
\centering
\caption{DermX diagnostic performance with regard to the gold standard. }

\begin{tabular}{ l r r r r } 
\toprule
   & F1 score & Sensitivity & Specificity \\
\midrule
Acne & $0.87 \pm 0.05$ & $0.89 \pm 0.07$ & $0.96 \pm 0.01$ \\
Actinic keratosis & $0.79 \pm 0.14$ & $0.74 \pm 0.14$ & $0.97 \pm 0.03$ \\
Psoriasis & $0.73 \pm 0.11$ & $0.77 \pm 0.11$ & $0.92 \pm 0.05$ \\
Seborrheic dermatitis & $0.74 \pm 0.09$ & $0.76 \pm 0.10$ & $0.93 \pm 0.04$ \\
Viral warts & $0.76 \pm 0.11$ & $0.68 \pm 0.15$ & $0.99 \pm 0.01$ \\
Vitiligo & $0.83 \pm 0.10$ & $0.82 \pm 0.13$ & $0.97 \pm 0.03$ \\ 
\midrule
Mean & $0.79 \pm 0.04$ & $0.78 \pm 0.04$ & $0.96 \pm 0.01$ \\
\bottomrule
\end{tabular}

\label{table:baseline_diagnostic_performance_gt}
\end{table}

\begin{table}[h!]
\centering
\caption{DermX+ diagnosis performance with regard to the gold standard. }

\begin{tabular}{ l r r r r } 
\toprule
   & F1 score & Sensitivity & Specificity \\
\midrule
Acne & $0.86 \pm 0.06$ & $0.92 \pm 0.06$ & $0.94 \pm 0.04$ \\ 
Actinic keratosis & $0.73 \pm 0.10$ & $0.69 \pm 0.15$ & $0.96 \pm 0.02$ \\ 
Psoriasis & $0.80 \pm 0.09$ & $0.83 \pm 0.08$ & $0.95 \pm 0.04$ \\ 
Seborrheic dermatitis & $0.74 \pm 0.10$ & $0.76 \pm 0.11$ & $0.94 \pm 0.03$ \\ 
Viral warts & $0.76 \pm 0.15$ & $0.70 \pm 0.18$ & $0.98 \pm 0.01$ \\ 
Vitiligo & $0.86 \pm 0.08$ & $0.83 \pm 0.11$ & $0.98 \pm 0.02$ \\ 
\midrule
Mean & $0.79 \pm 0.04$ & $0.79 \pm 0.04$ & $0.96 \pm 0.01$ \\
\bottomrule
\end{tabular}

\label{table:guided_attention_diagnostic_performance_gt}
\end{table}

\begin{table}[h!]
\centering
\caption{DermX performance on the presence or absence of characteristics with regard to the dermatologist-generated labels.}

\begin{tabular}{ l r r r r}
 \toprule
   & F1 score & Sensitivity & Specificity & Samples \\
 \midrule
Closed comedo & $0.76 \pm 0.08$ & $0.79 \pm 0.01$ & $0.95 \pm 0.03$ & 96\\
Dermatoglyph disruption & $0.74 \pm 0.20$ & $0.68 \pm 0.21$ & $0.99 \pm 0.02$ & 54 \\
Open comedo & $0.80 \pm 0.06$ & $0.82 \pm 0.09$ & $0.95 \pm 0.02$ & 110\\
Papule & $0.79 \pm 0.07$ & $0.79 \pm 0.10$ & $0.80 \pm 0.08$ & 278 \\
Patch & $0.76 \pm 0.06$ & $0.76 \pm 0.11$ & $0.82 \pm 0.06$ & 249\\
Plaque & $0.88 \pm 0.03$ & $0.89 \pm 0.03$ & $0.74 \pm 0.11$ & 352\\
Pustule & $0.79 \pm 0.10$ & $0.83 \pm 0.14$ & $0.94 \pm 0.02$ & 106\\
Scale & $0.79 \pm 0.04$ & $0.81 \pm 0.07$ & $0.77 \pm 0.07$ & 275\\
Scar & $0.78 \pm 0.10$ & $0.80 \pm 0.14$ & $0.94 \pm 0.04$ & 115\\
Sun damage & $0.66 \pm 0.11$ & $0.64 \pm 0.15$ & $0.96 \pm 0.03$ & 78\\
\midrule
Mean & $0.77 \pm 0.03$ & $0.78 \pm 0.04$ & $0.88 \pm 0.02$ & $171.30$\\
 \bottomrule
\end{tabular}
\label{table:baseline_characteristics_performance_binary}
\end{table}

\begin{table}[h!]
\centering
\caption{DermX+ performance on the presence or absence of characteristics with regard to the dermatologist-generated labels.}

\begin{tabular}{ l r r r r}
 \toprule
   & F1 score & Sensitivity & Specificity & Samples \\
 \midrule
Closed comedo & $0.81 \pm 0.12$ & $0.87 \pm 0.13$ & $0.94 \pm 0.04$ & 96\\
Dermatoglyph disruption & $0.70 \pm 0.2$ & $0.64 \pm 0.23$ & $0.99 \pm 0.01$ & 54 \\
Open comedo & $0.80 \pm 0.08$ & $0.85 \pm 0.09$ & $0.93 \pm 0.05$ & 110 \\
Papule & $0.80 \pm 0.07$ & $0.83 \pm 0.07$ & $0.76 \pm 0.11$ & 278 \\
Patch & $0.79 \pm 0.06$ & $0.77 \pm 0.11$ & $0.87 \pm 0.08$ & 249 \\
Plaque & $0.90 \pm 0.03$ & $0.89 \pm 0.05$ & $0.85 \pm 0.06$ & 352 \\
Pustule & $0.81 \pm 0.06$ & $0.86 \pm 0.08$ & $0.94 \pm 0.03$ & 106 \\
Scale & $0.82 \pm 0.05$ & $0.82 \pm 0.07$ & $0.83 \pm 0.07$ & 275 \\
Scar & $0.80 \pm 0.08$ & $0.82 \pm 0.14$ & $0.94 \pm 0.03$ & 115 \\
Sun damage & $0.64 \pm 0.15$ & $0.60 \pm 0.17$ & $0.96 \pm 0.03$ & 78 \\
\midrule
Mean & $0.79 \pm 0.03$ & $0.80 \pm 0.04$ & $0.90 \pm 0.02$ & $171.30$\\
 \bottomrule
\end{tabular}
\label{table:guided_attention_characteristics_performance_binary}
\end{table}

\begin{table}[h!]
\centering
\caption{DermX localization performance for localizable characteristics~(mean$\pm$std) with regard to the fuzzy dermatologist attention maps.}

\begin{tabular}{ l r r r r}
 \toprule
   & F1 score & Sensitivity & Specificity & Samples\\
 \midrule
Closed comedo & $0.40 \pm 0.04$ & $0.69 \pm 0.09$ & $0.69 \pm 0.05$ & 75 \\
Dermatoglyph disruption & $0.28 \pm 0.06$ & $0.69 \pm 0.09$ & $0.69 \pm 0.05$ & 36 \\
Open comedo & $0.36 \pm 0.05$ & $0.69 \pm 0.06$ & $0.68 \pm 0.04$ & 90 \\
Papule & $0.34 \pm 0.04$ & $0.63 \pm 0.08$ & $0.72 \pm 0.05$ & 219 \\
Patch & $0.43 \pm 0.05$ & $0.57 \pm 0.07$ & $0.78 \pm 0.04$ & 188 \\
Plaque & $0.43 \pm 0.03$ & $0.65 \pm 0.06$ & $0.75 \pm 0.04$ & 314 \\
Pustule & $0.24 \pm 0.05$ & $0.69 \pm 0.07$ & $0.69 \pm 0.05$ & 88 \\
Scale & $0.41 \pm 0.03$ & $0.65 \pm 0.07$ & $0.76 \pm 0.04$ & 222 \\
Scar & $0.46 \pm 0.05$ & $0.64 \pm 0.06$ & $0.72 \pm 0.05$ & 90 \\
Sun damage & $0.56 \pm 0.06$ & $0.44 \pm 0.08$ & $0.87 \pm 0.04$ & 50 \\
\midrule
Mean & $0.39 \pm 0.02$ & $0.64 \pm 0.02$ & $0.74 \pm 0.01$ & $137.20$\\
\bottomrule

\end{tabular}
\label{table:baseline_characteristics_performance_localisable_extra}

\end{table}

\begin{table}[h!]
    \centering
    \caption{DermX characteristics localization performance with regard to fuzzy dermatologist attention maps. The results include the localization performance of characteristics identified by the dermatologists but not by the model.}
    \begin{tabular}{l r r r r r}
        \toprule
         & F1 score & Sensitivity & Specificity & Samples \\
        \midrule 
        Closed comedo & $0.31 \pm 0.04$ & $0.55 \pm 0.11$ & $0.76 \pm 0.05$ & 96 \\
        Dermatoglyph disruption & $0.19 \pm 0.06$ & $0.48 \pm 0.17$ & $0.79 \pm 0.07$ & 54 \\
        Open comedo & $0.30 \pm 0.05$ & $0.57 \pm 0.08$ & $0.73 \pm 0.05$ & 110 \\
        Papule & $0.27 \pm 0.05$ & $0.50 \pm 0.08$ & $0.78 \pm 0.05$ & 278 \\
        Patch & $0.33 \pm 0.05$ & $0.44 \pm 0.08$ & $0.83 \pm 0.04$ & 249 \\
        Plaque & $0.38 \pm 0.02$ & $0.58 \pm 0.05$ & $0.77 \pm 0.04$ & 352 \\
        Pustule & $0.20 \pm 0.05$ & $0.58 \pm 0.14$ & $0.74 \pm 0.07$ & 106 \\
        Scale & $0.33 \pm 0.03$ & $0.53 \pm 0.08$ & $0.81 \pm 0.04$ & 275 \\
        Scar & $0.36 \pm 0.06$ & $0.50 \pm 0.08$ & $0.77 \pm 0.06$ & 115 \\
        Sun damage & $0.36 \pm 0.09$ & $0.28 \pm 0.07$ & $0.92 \pm 0.03$ & 78 \\
        \midrule
        Mean & $0.30 \pm 0.01$ & $0.50 \pm 0.01$ & $0.79 \pm 0.00$ & $171.30$ \\
        \bottomrule
    \end{tabular}
\label{table:baseline_characteristics_performance_localisable_extra_wrong}
\end{table}

\begin{table}[h!]
    \centering
    \caption{DermX+ localization performance for localizable characteristics~(mean$\pm$std) with regard to the fuzzy dermatologist attention maps.}
    
    \begin{tabular}{ l r r r r}
        \toprule
         & F1 score & Sensitivity & Specificity & Samples\\
        \midrule
        Closed comedo	& $0.10 \pm 0.09$ & 	$0.11 \pm 0.08$ &	$0.96 \pm 0.02$ & 83 \\	
        Dermatoglyph disruption	& $0.60 \pm 0.12$ & 	$0.60 \pm 0.15$ &	$0.97 \pm 0.02$ & 34 \\	
        Open comedo	& $0.06 \pm 0.05$ & 	$0.06 \pm 0.05$ &	$0.97 \pm 0.02$ & 93 \\	
        Papule	& $0.18 \pm 0.04$ & 	$0.19 \pm 0.06$ &	$0.97 \pm 0.02$ & 232 \\	
        Patch	& $0.53 \pm 0.05$ & 	$0.56 \pm 0.06$ &	$0.89 \pm 0.03$ & 191 \\	
        Plaque	& $0.61 \pm 0.05$ & 	$0.65 \pm 0.06$ &	$0.91 \pm 0.01$ & 312 \\	
        Pustule	& $0.03 \pm 0.03$ & 	$0.04 \pm 0.05$ &	$0.98 \pm 0.01$ & 91 \\	
        Scale	& $0.49 \pm 0.07$ & 	$0.58 \pm 0.11$ &	$0.89 \pm 0.03$ & 224 \\	
        Scar	& $0.35 \pm 0.12$ & 	$0.35 \pm 0.15$ &	$0.92 \pm 0.04$ & 93 \\	
        Sun damage	& $0.58 \pm 0.20$ & 	$0.53 \pm 0.19$ &	$0.90 \pm 0.10$ & 47 \\
        \midrule
        Mean & $0.36 \pm 0.01$ & $0.37 \pm 0.01$ & $0.94 \pm 0.00$ & $138.00$ \\
        \bottomrule
    
    \end{tabular}
\label{table:guided_attention_characteristics_performance_localisable_extra}

\end{table}

\begin{table}[h!]
    \centering
    \caption{DermX+ characteristics localization performance with regard to fuzzy dermatologist attention maps. The results include the localization performance of characteristics identified by the dermatologists but not by the model.}
    \begin{tabular}{l r r r r r}
        \toprule
         & F1 score & Sensitivity & Specificity & Samples \\
        \midrule 
        Closed comedo & $0.09 \pm 0.08$ & $0.09 \pm 0.07$ & $0.97 \pm 0.02$ & 96 \\
        Dermatoglyph disruption & $0.39 \pm 0.17$ & $0.4 \pm 0.21$ & $0.98 \pm 0.02$ & 54 \\
        Open comedo & $0.05 \pm 0.04$ & $0.05 \pm 0.04$ & $0.98 \pm 0.02$ & 110 \\
        Papule & $0.15 \pm 0.04$ & $0.16 \pm 0.06$ & $0.97 \pm 0.02$ & 278 \\
        Patch & $0.41 \pm 0.07$ & $0.43 \pm 0.07$ & $0.92 \pm 0.02$ & 249 \\
        Plaque & $0.54 \pm 0.06$ & $0.58 \pm 0.07$ & $0.92 \pm 0.01$ & 352\\
        Pustule & $0.03 \pm 0.03$ & $0.03 \pm 0.04$ & $0.98 \pm 0.01$ & 106 \\
        Scale & $0.41 \pm 0.07$ & $0.48 \pm 0.09$ & $0.91 \pm 0.02$ & 275 \\
        Scar & $0.29 \pm 0.1$ & $0.29 \pm 0.13$ & $0.93 \pm 0.04$ & 115 \\
        Sun damage & $0.34 \pm 0.14$ & $0.31 \pm 0.15$ & $0.94 \pm 0.08$ & 78 \\
        \midrule
        Mean & $0.27 \pm 0.01$ & $0.28 \pm 0.01$ & $0.95 \pm 0.00$ & $171.30$ \\
        \bottomrule
    \end{tabular}
\label{table:guided_attention_characteristics_performance_localisable_extra_wrong}
\end{table}


\end{document}